# Machine Learning Research Trends in Africa: A 30 Years Overview with Bibliometric Analysis Review

Absalom E. Ezugwu[1]*, Olaide N. Oyelade[2], Abiodun M. Ikotun[1], Jeffery O. Agushaka[1], Yuh-Shan Ho[3]*

[1]Unit for Data Science and Computing, North-West University, 11 Hoffman Street, Potchefstroom, 2520, South Africa

[2]Department of Computer Science, Faculty of Physical Sciences, Ahmadu Bello University, Zaria-Nigeria

[3]Trend Research Centre, Asia University, No. 500, Lioufeng Road, Wufeng, Taichung 41354, Taiwan

Authors E-mail: Absalom E Ezugwu (absalom.ezugwu@nwu.ac.za); Olaide N Oyelade (olaide_oyelade@yahoo.com); Abiodun M Ikotun (biodunikotun@gmail.com); Jeffrey O Agushaka (jefshak@gmail.com); Yuh-Shan Ho (ysho@asia.edu.tw)

* Corresponding Authors

**Draft: Due to copyediting, the published version is slightly different from this pre-print version**

**Abstract**. The machine learning (ML) paradigm has gained much popularity today. Its algorithmic models are employed in every field, such as natural language processing, pattern recognition, object detection, image recognition, earth observation and many other research areas. In fact, machine learning technologies and their inevitable impact suffice in many technological transformation agendas currently being propagated by many nations, for which the already yielded benefits are outstanding. From a regional perspective, several studies have shown that machine learning technology can help address some of Africa's most pervasive problems, such as poverty alleviation, improving education, delivering quality healthcare services, and addressing sustainability challenges like food security and climate change. In this state-of-the-art paper, a critical bibliometric analysis study is conducted, coupled with an extensive literature survey on recent developments and associated applications in machine learning research with a perspective on Africa. The presented bibliometric analysis study consists of 2761 machine learning-related documents, of which 89% were articles with at least 482 citations published in 903 journals during the past three decades. Furthermore, the collated documents were retrieved from the Science Citation Index EXPANDED, comprising research publications from 54 African countries between 1993 and 2021. The bibliometric study shows the visualization of the current landscape and future trends in machine learning research and its application to facilitate future collaborative research and knowledge exchange among authors from different research institutions scattered across the African continent.

*Keywords*: Machine learning; bibliometric study; web of science; Africa

**Abbreviations**

CU, Egypt: Cairo University, Egypt

UKZN, South Africa: University of KwaZulu-Natal, South Africa

UCT, South Africa: University of Cape Town, South Africa

MansU, Egypt: Mansoura University, Egypt

ZU, Egypt: Zagazig University, Egypt

UW, South Africa: University of Witwatersrand, South Africa

BU, Egypt: Benha University, Egypt

## 1. Introduction

The evolvement of the development and use of computers in intelligently solving problems predates the creation and testing of the Turing machine in 1950. Such systems aim to demonstrate their suitability in interfacing with human beings in a manner that shows a high level of intelligence compared to humans. However, this new set of systems was earlier motivated by the design of 1940s systems such as ENIAC, which aimed to emulate humans in promoting learning and thinking. The outcome of this led to computer game applications competitively gaming with humans. Furthermore, this motivated the design of perceptron, which accumulated into a broader design of machine learning used for classification purposes. Further research and applications in statistics have promoted machine learning so that the intersection of statistics and computer science has advanced studies on artificial intelligence (AI). In this section, we organize the discussion to provide background knowledge on AI, ML and deep learning (DL). We provide a summary of a multi-disciplinary approach to research on ML to show recent methods and major application areas of ML in addressing real-problems. We conclude this section by providing a motivation for the bibliometric analysis and highlighting the study's contribution.

**1.1 A brief background of ML and its evolution from AI-ML-DL**

The drive to replace human capability with machine intelligence led to the evolvement of various methods of AI, which is now defined as the science and engineering of achieving machine intelligence as often exhibited in the form of computer programs and often in controlling and receiving signals from hardware (Cioffi et al., 2020). An upsurge of research in AI has resulted in the outstanding performance of machines that now perform complex tasks intelligibly. Several AI paradigms have now evolved, including natural language processing (NLP), constraint satisfaction, machine learning, distributed AI, machine reasoning, data mining, expert systems, case-based reasoning (CBR), knowledge-representation, programming, robotics, belief revision, neural network, theorem proving, theory computation, logic, and genetic algorithm. This evolvement follows a historical trend, as shown in Figure 1, which demonstrates a continuous improvement of methods and algorithms to increase accuracy in the exhibition of machine intelligence. This timeline shows that research in AI advanced through some challenging exploits until around the 1970s, when ML conceptualization began to manifest interesting results and performances. Interestingly, with these advances came the challenge of addressing ethical issues so that AI-driven systems are not allowed to infringe on human rights, nor will the moral status of such systems be compromised (Bostrom N & Yudkowsky, 2010). That notwithstanding, the evolvement peaked from the basic Turing's concept to the current Industry 4.0 by connecting multi-disciplinary approaches, including those from computer science but also psychology, philosophy, neuroscience, biology, mathematics, sociology, linguistics, and other areas (Amudha, 2021).

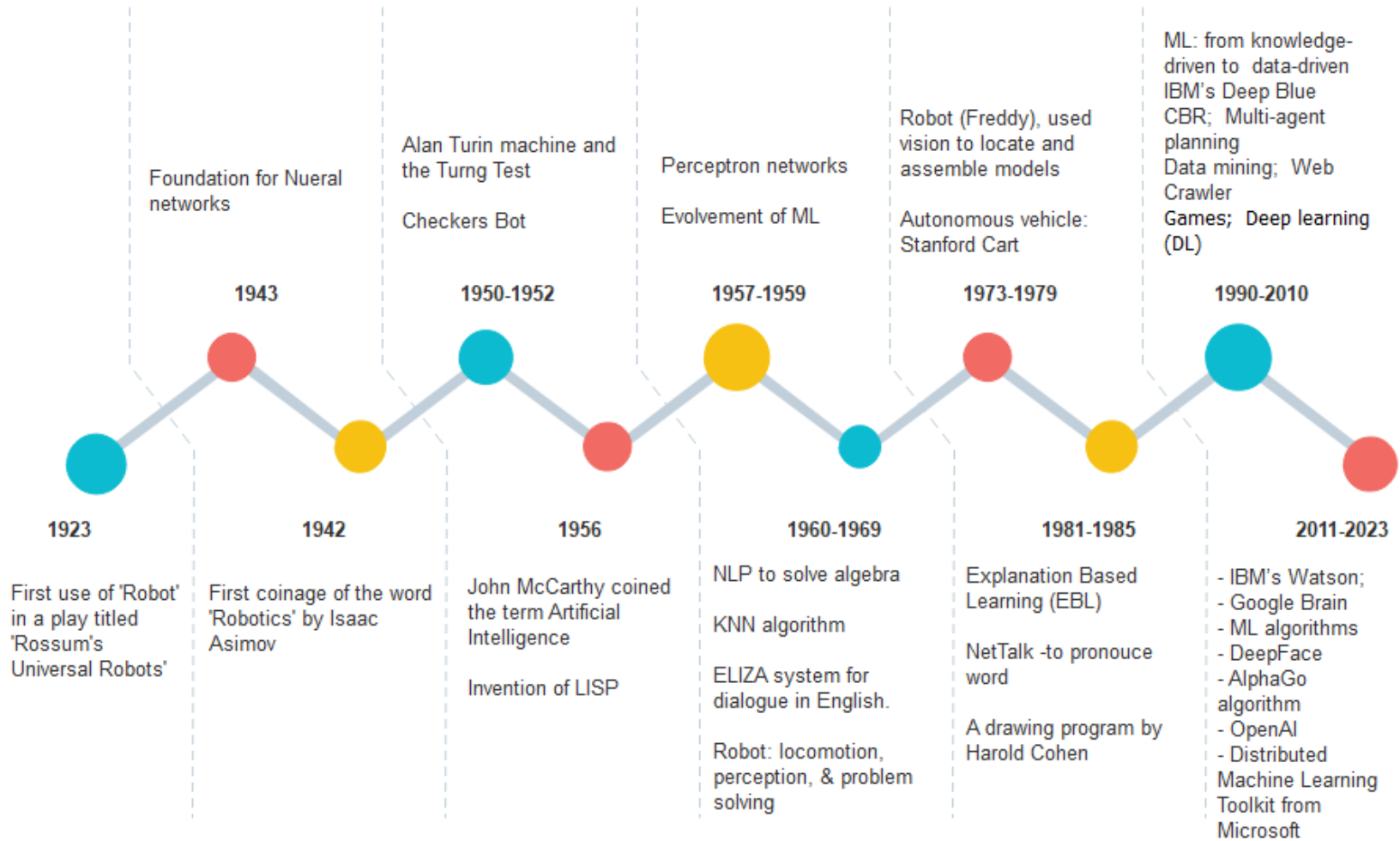

Figure 1: Evolvement of AI from dream to reality

The field of machine learning (ML) branched out of AI and is focused on evolving computational methods and algorithms learning and building learning machines to leverage an object's natural pattern of learning features. ML has been reputed to advance AI dramatically because of its problem-solving approach of recognizing patterns in domain-specific datasets to gather artificial experience from the observed data. This follows a data extraction pipeline through training and prediction using new data. This learning, shown in Figure 2, is approached from and has evolved into different perspectives, including popular supervised, unsupervised, semi-supervised, and reinforcement learning. Over the years, algorithms have been designed and further evolved in each aspect of learning. These algorithms address real-life problems involving classification and regression problems using supervised learning methods, clustering and association using unsupervised learning methods, and the problem of understanding and manoeuvring an environment using reinforcement learning. The learning process in ML uses both symbolic and numeric methods as incorporated into some of its popular algorithms such as linear regression, nearest neighbor, Gaussian Naive Bayes, decision trees, support vector machine (SVM), random forest, K-Means, density-based spatial clustering of applications with noise (DBSCAN), balanced iterative reducing and clustering (BIRCH), temporal difference (TD), Q-Learning, and deep adversarial networks. The design of these algorithms includes a broad domain of statistics, genetic algorithms, computational learning theory, neural networks, stochastic modeling, and pattern recognition. The resulting algorithms have demonstrated state-of-the-art performances in email filters, NLP, pattern recognition, computer vision and autonomous vehicle design.

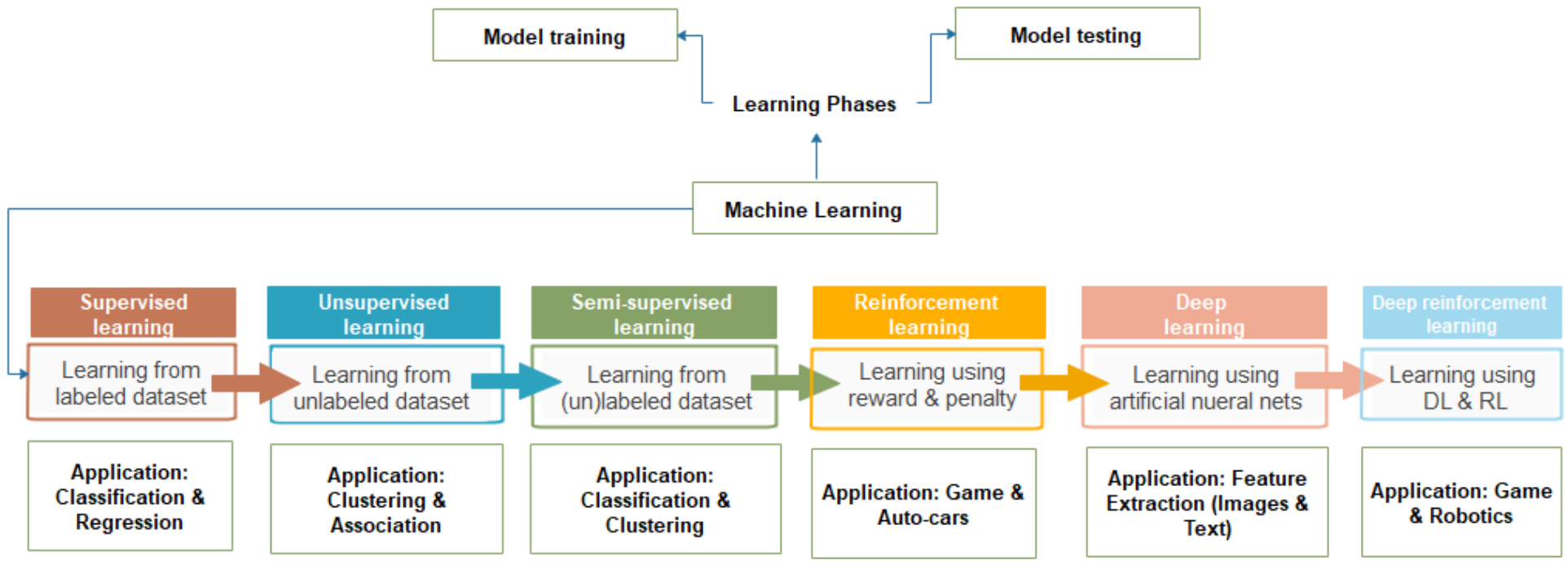

Figure 2: Evolvement of machine learning evolved from supervised learning to reinforcement learning

Deep learning (DL) belongs to the broader family of ML and can analyse data intelligently through transformations, graph technologies and representation patterns. Derived from the simulation of the human brain from the basic Artificial Neural Networks (ANN), convolutional neural networks are designed in a manner that outperforms traditional ML algorithms. The approach leverages increasingly available training data from sensors, the Internet of Things (IoT), surveillance systems, intrusion detection system, cybersecurity, mobile, business, social media, health, and other devices. These data, often in an unstructured format, are analyzed and automated for identification of features leading to either classification or regression analysis (Kolajo et al., 2019). The DL has been widely adapted to address application problems, including audio and speech, visual data, and NLP. Design patterns for DL have appeared as Convolutional neural network (CNN) - the most popular and widely used of DL networks - recursive neural networks (RvNNs), recurrent neural networks (RNNs), Boltzmann machine (BM), and auto-encoders (AE). While the RNN is often applied to text or signal processing, RvNN, which uses a hierarchical structure, can classify outputs utilizing compositional vectors (Alzubaidi et al., 2021). Results obtained from different studies showed that DL had obtained good outstanding performance across a variety of applications (Oyelade & Ezugwu, 2020). This has now motivated its integration into reinforcement learning to achieve Deep Reinforcement Learning (DRL). Considering this evolvement and performance of ML and DL, we focus the next sub-section on presenting brief research and application of methods in this field among African researchers.

**1.2 A brief background of multi-disciplinary ML research contribution from different scientists across major African universities**

There is widespread research using ML to address contextual problems across African countries. In most cases, this is promoted by a local conference called Indaba, which promotes the application of DL and ML to help ensure that knowledge, capacity, recognizing excellence in ML research, and application are well harnessed to develop the continent. In this section, a summary of studies on ML and DL in Africa is reviewed to demonstrate the level of involvement of the researchers in research on ML. It is reported that AI-based research is improving communities across the Sub-Saharan Africa (SSA) regions. In Kenya, it is being applied to aid health worker–patient interaction to detect blinding eye disorders, and in Egypt, in aiding automated decision-making systems for health-care support. In South Africa, it is aiding drug prescription, and with a multinomial logistic classifier-based application, it is being applied to human resource planning. ML-trained models are primarily deployed in medicine in Nigeria, and an example is their use in the diagnosis of birth asphyxia and identification of fake drugs. Other cases are the use of ML to diagnose diabetic retinopathy in Zambia and the diagnosis of pulmonary tuberculosis in Tanzania (Owoyemi et al., 2020).

Studies in the ML application from Morocco cut across medicine, solar power and climate. In particular, deep learning models, CNN, have been proposed for detecting and classifying breast cancer cases using histopathology samples (Agouri et al., 2022). The RNN variant of a DL model has been adapted to address the problem of daily streamflow over the Ait Ouchene watershed (AIO). The study used the Short-Term Long Memory (LSTM) network, a type of RNN, to achieve this simulation (Nifa et al., 2023). Research applying ML methods in the remote sensing field using a popular algorithm such as support vector machines (SVM) in mapping Souk Arbaa Sahel in a lithological manner has been reported by Bachri et al. (2019). In the financial sector, researchers have investigated the use of ML in revolutionizing the banking ecosystem for precise credit scoring, regulation and operational approaches (Hamdoun & Rguibi, 2019). In another study, the country's location motivates research on using ML to harness solar power in grid management at power plants. Both ML algorithms and DL have been drafted for predicting solar radiation using models such as ANN, multi-layer perceptron (MLP), back propagation neural network (BPNN), deep neural network (DNN), and LSTM (Boutahir et al., 2022).

In Egypt, research in ML has enjoyed application to learner-ship, face recognition, visual surveillance, and optical character recognition (OCR). In a study, the rate of school-dropout has been investigated and predicted using ML algorithms, specifically a Logistic classifier. The model can identify students at-risk of dropping out of school and isolate the causative of this challenge (Selim & Rezk, 2023). A novel hybrid DL model capable of detecting features supportive of face recognition has been proposed to apply the trained model to build a face clustering system based on density-based spatial clustering of applications with noise (DBSCAN) (Ahmed et al., 2020). Similarly, generative adversarial networks (GANs), a composition of DL models adversarial positioned for generative purposes, have been investigated for kinship face synthesis (Ghatas & Hemayed, 2020). Also, identification systems have been built using CNN by extracting input from video files to apply vision surveillance (Bayoumi et al., 2022). The contextualization of optical character recognition (OCR) systems to solve local problems has been researched using CNN, DNN and the SVM classifier to recognise different classes accurately (Sokar et al., 2018). Another interesting application of DL is in the task of Automatic License Plate Detection and Recognition (ALPR) for Egyptian license plates (ELP) (Elnashar et al., 2020).

The ML and DL models have been used primarily in Nigeria's medicine, security and climate issues. For instance, the use of CNN in investigating a solution to the classification problem of breast cancer using digital mammograms has been reported (Oyelade & Ezugwu, 2022). In related work, performance enhancement techniques such as data augmentation in improving DL models have been researched (Oyelade & Ezugwu, 2021) using the CNN model to detect architectural distortion in breast images. Concerning the challenge of deploying ML methods to address COVID-19, studies have been conducted using DL architectures to detect and classify the disease in chest x-ray samples (Oyelade et al., 2021). Similarly, the need to harness the deployment of Internet of Things (IoT) devices to curb the spread of COVID-19 using ML algorithms has been advocated (Ezugwu et al., 2021). On the issue of security, an investigative study has been carried out assessing the level of deployment of AI and its associated ML methods in curbing terrorism and insurgency in Nigeria (NSUDE, 2022). The use of artificial neural network (ANN) and logistic regression (LR) models have also been used to predict floods in susceptible areas in Nigeria (Ighile et al., 2022). Regarding finance and the digital economy, AI-based methods have been recommended for innovation and policy-making (Robinson, 2018).

Researchers in Uganda have also employed AI in healthcare management by observing the performance of an AI algorithm called Skin Image Search, applied to dermatological tasks. The algorithm was trained using a local dataset from The Medical Concierge Group (TMCG) to diagnostically analyze and extract the gender, age and dermatological diagnosis (Kamulegeya et al., 2019). A researcher from Kenya confirmed that an investment of US$74.5 million is being made to support the use of ML models in healthcare (Waljee et al., 2022). In the same country, DL architecture, namely the LSTM network, has been investigated for drought management by forecasting vegetation's health (Lees et al., 2022).

Research on the application of ML is widespread in South Africa, with more consideration given to language processing, medical image analysis, and astronomy. In addition to using ML algorithms, DL and NLP have been well-researched to aid development (Biljon, 2022). Generative model GAN has been applied to enable automatic speech recognition (ASR), improving the features of mismatched data prior to decoding (Heymans, Davel, & Heerden, 2022). In another related work, the ASR system has been researched by combining multi-style training (MTR) with deep neural network hidden Markov model (DNN-HMM) (Heymans, Davel, & van Heerden, 2022). The use of CNN in exploring classification accuracy on SNR data has been reported by Andrew et al. (2021). A study has been channeled to investigate the role of loss functions in aiding the behavior of deep neural network optimization purposes (Venter et al., 2020). Feedforward neural networks have been used to study the space physics problem in storm forecasting (Beukes et al., 2020). Optimizing hyperparameter issues in embedding algorithms has been considered for improving training word embeddings with speech-recognized data (Barnard & Heyns, 2020).

All these clear indications show that there is now a strong increase in research in ML, including its associated sub-fields of DL and NLP in African universities, with most applications aimed at healthcare, climate, and security. In the following sub-section, we summarize the major application areas of ML in the continent. This is necessary to give perspective to the current state of research on ML in the domain and to serve as a motivation for enabling future research on ML.

**1.3 A brief highlight on the significance of ML application within the continent**

Findings from the reviewed process detailed in the study showed that the fields of medicine and healthcare delivery management, agricultural studies, security and surveillance, natural language modelling and process and many others had benefited immensely from the application of ML on the continent. These ML applications include research on DL in cyber security intrusion detection and, likewise, the detection of DDoS in cloud computing. Disease detection in plants and crops has also been investigated using ML algorithms with an example of tomato disease detection. The sugarcane leaf nitrogen concentration estimation has been reported to map irrigated areas using Google Earth Engine. Several sub-fields of medicine have received research attention in promoting healthcare delivery and improving disease detection and management. Examples of ML methods in this aspect are automatic sleep stage classification, face mask detection in the era of the COVID-19 pandemic, protein sequence classification, and temporal gene expression data. Several studies have also been applied to study the design of optimization and clustering methods to solve difficult optimization problems in engineering, medicine and science. NLP methods have received wider consideration and study for mainstreaming the use of local languages across the continent. This includes translating the Yoruba language to French, automation regarding the use of Swahili, and automatic Arabic Diacritization. Other interesting areas generating the application of ML algorithms on the continent are optical communications and networking (Musumeci et al., 2018), deployment of AI to software engineering problems (Batarseh et al., 2020), and advancing medical research and appropriating clinical artificial intelligence in check-listing research (Olczak et al., 2021).

**1.4 Strong motivation and need for the current employment of bibliometric analysis study**

This study is motivated by the availability of large research databases providing a considerable number of publications and research outputs suitable for aiding the search required for the study. This data availability has helped to guide the decision on the need to use bibliometric techniques in drawing out important findings from the data collected from the scientific databases. Bibliometrics is used to facilitate the examination of large bodies of knowledge within and across disciplines. The use of bibliometric techniques in this study will support the aim of the study in identifying hidden but useful patterns capable of illustrating the research trend on ML and DL by researchers in African universities. This study intends to leverage the presentational nature of bibliometric analysis to allow policymakers to easily discover interesting research works in ML on the continent to aid their decision-making process.

Interestingly, we found that the proposed method will allow for discovering leading contributors to ML research. This method will undoubtedly enable this study to uncover new directions and themes for future research in ML. As observed in subsequent sections, bibliometric analysis enabled us to evaluate the impact of publications by regions, research institutions and authors and obtain relevant scientific information on a topic. The quantitative, scalable and transparent approach of bibliometric analysis fits them closely as informetrics and scientometrics. In the next sub-section, the approach to applying the bibliometric techniques in this study to achieve the aim of the study is outlined.

The following highlights are the major contributions of this study:

- We first apply the analysis of research publications to uncover the developments with ML in African universities.

- The study identifies core research in ML and DL and authors and their relationship by covering all the publications from African researchers.

- We analyze the research status and frontier directions and predict the future of ML research in Africa.

- An analysis of entities such as authors, institutions or countries in African universities is compared to their research outputs.

The remaining part of the paper is organized as follows: Section 2 describes the data collection process and the methodology used in this paper. Extensive bibliometric analysis is performed in Section 3, and this section covers the presentation of significant narratives and a detailed discussion of findings from the conducted study analysis. We provide a detailed literature review of the last few years in Section 4. Section 5 concludes the paper by summarizing the study's findings of 30 years of ML-dedicated research efforts in several universities across the African continent.

**2. Methodology**

To do bibliometric analyses, data were extracted from the online databases of the Science Citation Index Expanded (SCI-EXPANDED) (data extracted on 10 October 2022). Quotation marks (" ") and Boolean operator "or" were used, which ensured the appearance of at least one search keyword in terms of TOPIC (title, abstract, author keywords, and *Keywords Plus*) from 1991 to 2021 (Al-Moraissi et al., 2022). The search keywords: "machine learning", "machining learning", "machine learnable", "machine learn", "machine learns", "machine learners", "machine learner", "machine learnings", "machines learning", "machine learnt", "machine learned", and "machines learn" that were found in SCI-EXPANDED were considered. To have accurate analysis results, some terms missed spaces were found and employed including "machine learningmethods", "machine learningmetrics", "machine learningbased", "machine learningalgorithm", and "machine learningclassifiers". Furthermore, related keywords which were misspelt such as "machine learnig", "machine learnin", "maching learning", and "machin learning" were also used as search keywords. African countries including "Algeria", "Angola", "Benin", "Botswana", "Burkina Faso", "Burundi", "Cameroon", "Cape Verde", "Cent Afr Republ", "Chad", "Comoros", "Dem Rep Congo", "Rep Congo", "Cote Ivoire", "Djibouti", "Egypt", "Equat Guinea", "Eritrea", "Eswatini", "Ethiopia", "Gabon", "Gambia", "Ghana", "Guinea", "Guinea Bissau", "Kenya", "Lesotho", "Liberia", "Libya", "Madagascar", "Malawi", "Mali", "Mauritania", "Mauritius", "Morocco", "Mozambique", "Namibia", "Niger", "Nigeria", "Rwanda", "Sao Tome & Prin", "Senegal", "Seychelles", "Sierra Leone", "Somalia", "South Africa", "South Sudan", "Sudan", "Tanzania", "Togo", "Tunisia", "Uganda", "Zambia", and "Zimbabwe" were also searched in terms of the country (CU). A total of 2,770 documents, including 2,477 articles, were found in SCI-EXPANDED from 1993 to 2021. In summary, a PRISMA flow diagram is shown in Figure 3. It visually depicts the review process of finding published data on the topic and the authors decisions' on whether to include it in the review. This study selected only articles from Science Citation Index Expanded (SCI-EXPANDED) with keywords as explained earlier. Quotation marks (" ") and Boolean operator "or" were used, which ensured the appearance of at least one search keyword in terms of TOPIC (title, abstract, author keywords, and Keywords Plus) from 1991 to 2021.

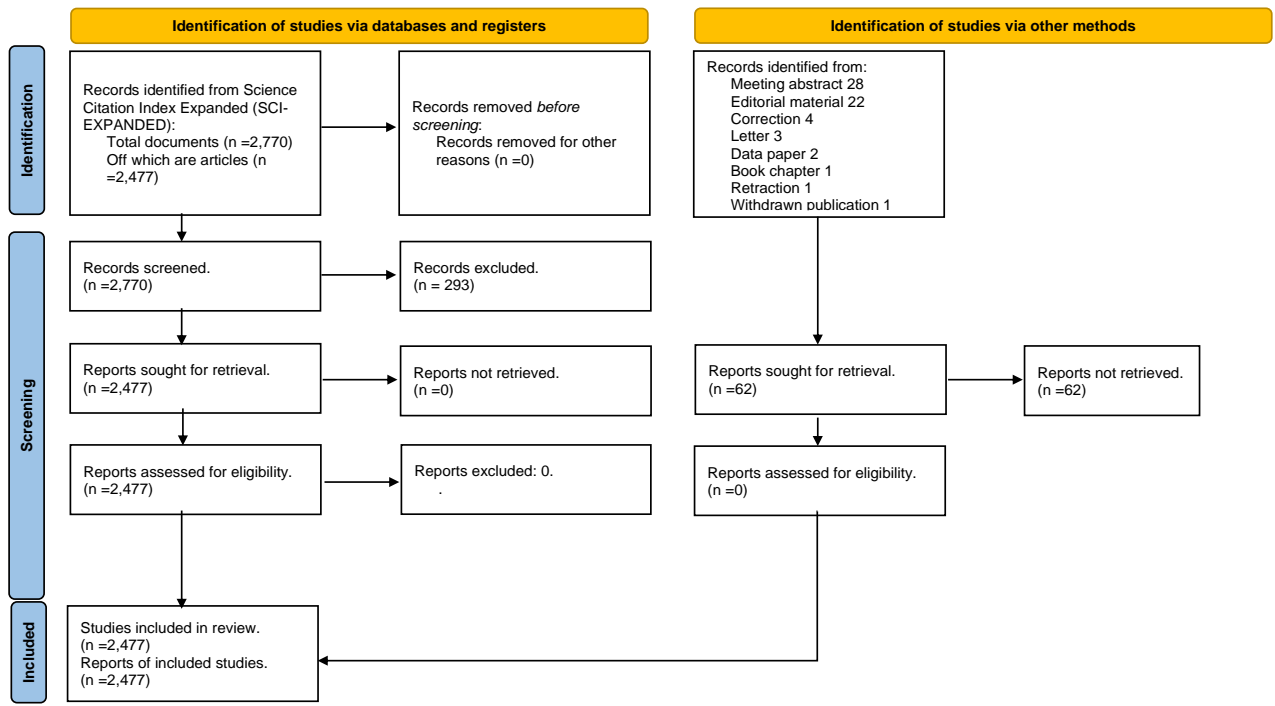

Figure 3: A summary of the data extraction and screening process from SCI-EXPANDED

*Keywords Plus* provides additional search terms extracted from the titles of articles cited by authors in their bibliographies and footnotes in the Institute of Science Information (ISI) (now Clarivate Analytics) database. It substantially augments title-word and author-keyword indexing (Garfield, 1990). It was noticed that documents only searched out by Keywords Plus are irrelevant to the search topic (Fu and Ho, 2015). Ho's group first proposed the "front page" as a filter to improve bias by using the data from SCI-EXPANDED directly, including the article title, abstract, and author keywords (Wang and Ho, 2011). It has been pointed out that a significant difference was found by using the 'front page' as a filter in bibliometric research in wide journals classified in SCI-EXPANDED, for example, *Frontiers in Pharmacology* (Ho, 2019a), *Chinese Medical Journal* (Ho, 2019b), *Environmental Science and Pollution Research* (Ho, 2020a), Water (Ho, 2020b), *Science of the Total Environment* (Ho, 2021), and *Journal of Foot and Ankle Surgery* (Ho, 2022). The 'front page' filter can avoid introducing unrelated publications for bibliometric analysis.

The entire record and the annual number of citations for each document were checked and placed into Excel Microsoft 365, and additional coding was manually executed. The functions in Excel Microsoft 365, for example, Concatenate, Counta, Freeze Panes, Len, Match, Proper, Rank, Replace, Sort, Sum, and Vlookup, were applied. The journal impact factors ($IF_{2021}$) were based on the Journal Citation Reports (JCR) issued in 2021.

In the SCI-EXPANDED database, the corresponding author is designated as the "reprint author"; "corresponding author" will continue to be the primary term rather than the reprinted author (Ho, 2012). In single-author articles where authorship is not specified, the single author is considered the first and corresponding author (Ho, 2014a). Likewise, in single-institutional articles, institutions are classified as first-author and corresponding author institutions (Ho, 2014a). All corresponding authors, institutions, and countries were considered in multiple corresponding author articles. For more accurate analysis results, affiliations were checked and reclassified. Author affiliations in England, Scotland, North Ireland (Northern Ireland), and Wales were regrouped under the heading of the United Kingdom (UK) (Chiu and Ho, 2007). Furthermore, SCI-EXPANDED has the article of the corresponding author. Only the address without the name of the affiliations is found, and the address is changed to the name of the affiliations.

Six publication indicators are used to assess the publication performance of countries and institutions (Hsu and Ho, 2014; Mohsen & Ho, 2022): *TP*: total number of articles; *IP*: number of single-country ($IP_C$) or single-institution articles ($IP_I$); *CP*: number of internationally collaborative articles ($CP_C$) or inter-institutionally collaborative articles ($CP_C$); *FP*: number of first-author articles; *RP*: number of corresponding-author articles; and *SP*: number of single-author articles. Moreover, publications were assessed using the following citation indicators: $C_{year}$: the number of citations from Web of Science Core Collection in a year (e.g. $C_{2021}$ describes citation count in 2021) (Ho, 2012); and $TC_{year}$: the total citations from Web of Science Core Collection received since publication year till the end of the most recent year (2021 in this study, $TC_{2021}$) (Wang et al., 2011; Mohsen & Ho, 2022).

Six citation indicators ($CPP_{2021}$) related to the six publication indicators were also applied to evaluate the publication's impact on countries and institutions (Ho and Mukul, 2021): $TP\text{-}CPP_{2021}$: the total $TC_{2021}$ of all articles per the total number of articles ($TP$); $IP\text{-}CPP_{2021}$: the total $TC_{2021}$ of all single-country articles per the number of single-country articles ($IP_C\text{-}CPP_{2021}$) or single-institutions articles per the number of single-institutions articles ($IP_I\text{-}CPP_{2021}$); $CP\text{-}CPP_{2021}$: the total $TC_{2021}$ of all internationally per the number of internationally collaborative articles ($CP_C\text{-}CPP_{2021}$) or inter-institutionally collaborative articles per inter-institutionally collaborative articles ($CP_I\text{-}CPP_{2021}$); $FP\text{-}CPP_{2021}$: the total $TC_{2021}$ of all first-author articles per the number of first-author articles ($FP$); $RP\text{-}CPP_{2021}$: the total $TC_{2021}$ of all corresponding-author articles per the number of corresponding-author articles ($RP$); and $SP\text{-}CPP_{2021}$: the total $TC_{2021}$ of all single-author articles per the number of single-author articles ($SP$).

### 3. Results and Discussion

### 3.1. Document type and language of publication

The characteristics of document type based on their $CPP_{\text{year}}$ and the average number of authors per publication ($APP$) as basic document type information in a research topic were proposed (Monge-Nájera and Ho, 2017). Recently, the median of the number of authors was also applied to a research topic with a large number of authors in a document (Elhassan et al., 2022). Using the citation indicators $TC_{\text{year}}$ and $CPP_{\text{year}}$ has advantages compared to citation counts directly from the Web of Science Core Collection because of their invariance and ensuring reproducibility (Ho and Hartley, 2016a). A total of 2,761 machine learning-related documents by authors affiliated with several institutions in Africa published in the SCI-EXPANDED from 1993 to 2021 were found among 11 document types which are detailed in Table 1. The majority were articles (89% of 2,761 articles) with an *APP* of 15 and a median of 4.0.

Table 1. Citations and authors based on the document types.

| Document type | TP | % | AU | APP | Median | $TC_{2021}$ | $CPP_{2021}$ |
|---|---|---|---|---|---|---|---|
| Article | 2,468 | 89 | 37,770 | 15 | 4.0 | 28,350 | 11 |
| Review | 235 | 8.5 | 1,391 | 5.9 | 4.0 | 4,287 | 18 |
| Proceedings paper | 32 | 1.2 | 213 | 6.7 | 3.0 | 365 | 11 |
| Meeting abstract | 28 | 1.0 | 359 | 13 | 7.5 | 15 | 0.54 |
| Editorial material | 22 | 0.80 | 113 | 5.1 | 3.0 | 88 | 4.0 |
| Correction | 4 | 0.14 | 24 | 6.0 | 6.0 | 0 | 0 |
| Letter | 3 | 0.11 | 24 | 8.0 | 4.0 | 49 | 16 |
| Data paper | 2 | 0.072 | 29 | 15 | 15 | 16 | 8.0 |
| Book chapter | 1 | 0.036 | 1 | 1.0 | 1.0 | 0 | 0 |
| Retraction | 1 | 0.036 | 4 | 4.0 | 4.0 | 0 | 0 |
| Withdrawn publication | 1 | 0.036 | 1 | 1.0 | 1.0 | 4 | 4.0 |

*TP*: total number of publications; *AU*: number of authors; *APP*: average number of authors per publication; $TC_{2021}$: total number of citations from Web of Science Core Collection since publication year to the end of 2021; $CPP_{2021}$: average number of citations per publication ($TC_{2021}/TP$).

The largest number of authors in an article is "Y Machine learning risk prediction of mortality for patients undergoing surgery with perioperative SARS-CoV-2: the COVIDSurg mortality score" (Bravo et al., 2021) published by 4,819 authors from 784 institutions in 71 countries including African countries: Egypt, Ethiopia, Gabon, Libya, Morocco, Nigeria, South Africa, Sudan, and Zimbabwe. The document type of reviews with 235 documents had the greatest $CPP_{2021}$ value of 18, which was 1.6 times of articles. Five of the top 12 most frequently cited documents were reviews by Carleo et al. (2019) ($TC_{2021}$ = 426; rank 3rd), Merow et al. (2014) ($TC_{2021}$ = 268; rank 6th), Nathan et al. (2012) ($TC_{2021}$ = 231; rank 9th), Oussous et al. (2018) ($TC_{2021}$ = 206; rank 11th), and Ben Taieb et al. (2012) ($TC_{2021}$ = 204; rank 12th).

Web of Science document type of articles were further analyzed as they included the entire research hypothesis, methods and results (Ho, 2010). Only three non-English articles were published by the French in *Traitement du Signal* (Bahi, 2007; Ballihi et al., 2012) and *Annales Des Télécommunications* (Nouali and Blache, 2005).

### 3.2. Characteristics of publication outputs

A relationship between the annual number of articles (*TP*) and their $CPP_{year}$ by the years in a research field has been applied as a unique indicator (Ho, 2013; Ho et al., 2022). Machine learning research was not considered in Africa before 2010, with an annual number of articles of less than 10. In Africa, S. Elgamal, M. Rafeh, and I. Eissa from Cairo University in Egypt first mentioned "machine learning" as the authors' keywords in *Case-based reasoning algorithms applied in a medical acquisition tool* (Elgamal et al., 1993). The number of articles increased slightly from 14 in 2010 to 98 in 2017 (Figure 4). After that, a sharply rising trend reached 1,035 articles in 2021. The highest $CPP_{2021}$ was 54 in 2013, which can be attributed to the article entitled *Multiobjective intelligent energy management for a microgrid* (Chaouachi et al., 2013), ranking at the top in $TC_{2021}$ with 402 (rank 3rd).

### 3.3. Web of Science categories and journals

African published machine learning-related articles in 903 journals were classified in 159 of the 178 Web of Science categories in SCI-EXPANDED. Recently, the characteristics of the Web of Science categories based on *TP*, *APP*, $CPP_{2021}$, and the number of journals in each category were proposed (Giannoudis et al., 2021). Table 2 shows the top 12 productive Web of Science categories with over 100 articles. A total of 906 articles (37% of 2,468 articles) were published in the top four productive categories: electrical and electronic engineering containing 278 journals (385 articles; 20% of 2,468 articles), information systems computer science containing 164 journals (439 articles; 18%), artificial intelligence computer science containing 145 journals (334 articles; 14%), and telecommunications containing 94 journals (308 articles; 12%). Comparing the top 12 productive categories, articles published in the 'interdisciplinary applications computer science' and 'remote sensing' categories had the greatest $CPP_{2021}$ of 15, respectively. Articles published in the 'information systems computer science category' had a lower $CPP_{2021}$ of 8.9. Articles published in the category of 'environmental sciences' had the greatest $APP$ of 6.6, while articles in the category of 'artificial intelligence computer science' had an $APP$ of 3.5. The interaction of publication development among Web of Science categories is discussed using Figure 4, comprising the number of publications versus the year of publication (Ho et al., 2010). Figure 5 shows the development trends of the top four Web of Science categories with more than 300 articles. The first articles were published in 1993 and 1997 in the 'information systems computer science' and 'electrical and electronic engineering' categories, respectively. However, more articles have been published in the 'electrical and electronic engineering' category since 2014. The first article in the category of 'telecommunications' was found in 2015. It had a sharp increase since 2018 and reached 139 articles in 2021, much higher than the 93 articles in the 'artificial intelligence computer science category'.

Table 2. Top 12 most productive Web of Science categories with $TP > 100$

| Web of Science category | $TP$ (%) | No. $J$ | APP | $CPP_{2021}$ |
| --- | --- | --- | --- | --- |
| electrical and electronic engineering | 485 (20) | 278 | 4.3 | 13 |
| information systems computer science | 439 (18) | 164 | 4.4 | 8.9 |
| artificial intelligence computer science | 334 (14) | 145 | 3.5 | 14 |
| Telecommunications | 308 (12) | 94 | 4.4 | 9.3 |
| environmental sciences | 192 (7.8) | 279 | 6.6 | 11 |
| multidisciplinary sciences | 163 (6.6) | 73 | 6.2 | 13 |
| interdisciplinary applications computer science | 154 (6.2) | 113 | 5.7 | 15 |
| multidisciplinary geosciences | 142 (5.8) | 202 | 5.9 | 12 |
| theory and methods computer science | 138 (5.6) | 110 | 4.3 | 11 |
| remote sensing | 120 (4.9) | 34 | 5.6 | 15 |
| imaging science and photographic technology | 108 (4.4) | 28 | 5.3 | 13 |
| energy and fuels | 101 (4.1) | 119 | 4.5 | 11 |

*TP*: total number of publications; *No. J*: number of journals in a category in 2021; *APP*: average number of authors per publication; $CPP_{2021}$: average number of citations per publication ($TC_{2021}/TP$).

Recently, the characteristics of the journals based on their $CPP_{year}$ and *APP* as basic information of the journals in a research topic were proposed (Ho, 2021; Al-Moraissi et al., 2022). Table 3 shows the top 12 most productive journals with journal impact factors, $CPP_{2021}$, and *APP*. The *IEEE Access* ($IF_{2021}$ = 3.476) published the most, 192 articles, representing 7.8% of 2,468. Compared to the top 12 productive journals, articles published in the *Expert Systems with Applications* ($IF_{2021}$ = 8.665) had the greatest $CPP_{2021}$ of 30. In contrast, articles in the *CMC-Computers Materials & Continua* ($IF_{2021}$ = 3.860) had only 2.2. The *APP* ranged from 16 in the *Monthly Notices of the Royal Astronomical Society* to 2.8 in the *Journal of Big Data*. According to $IF_{2021}$, the top five journals which have an $IF_{2021}$ of more than 60 were *World Psychiatry* ($IF_{2021}$ = 79.683) with two articles, *Nature* ($IF_{2021}$ = 69.504) with one article, *Nature Energy* ($IF_{2021}$ = 67.439) with one article, *Nature Reviews Disease Primers* ($IF_{2021}$ = 65.038) with one article, and *Science* with one article ($IF_{2021}$ = 63.714).

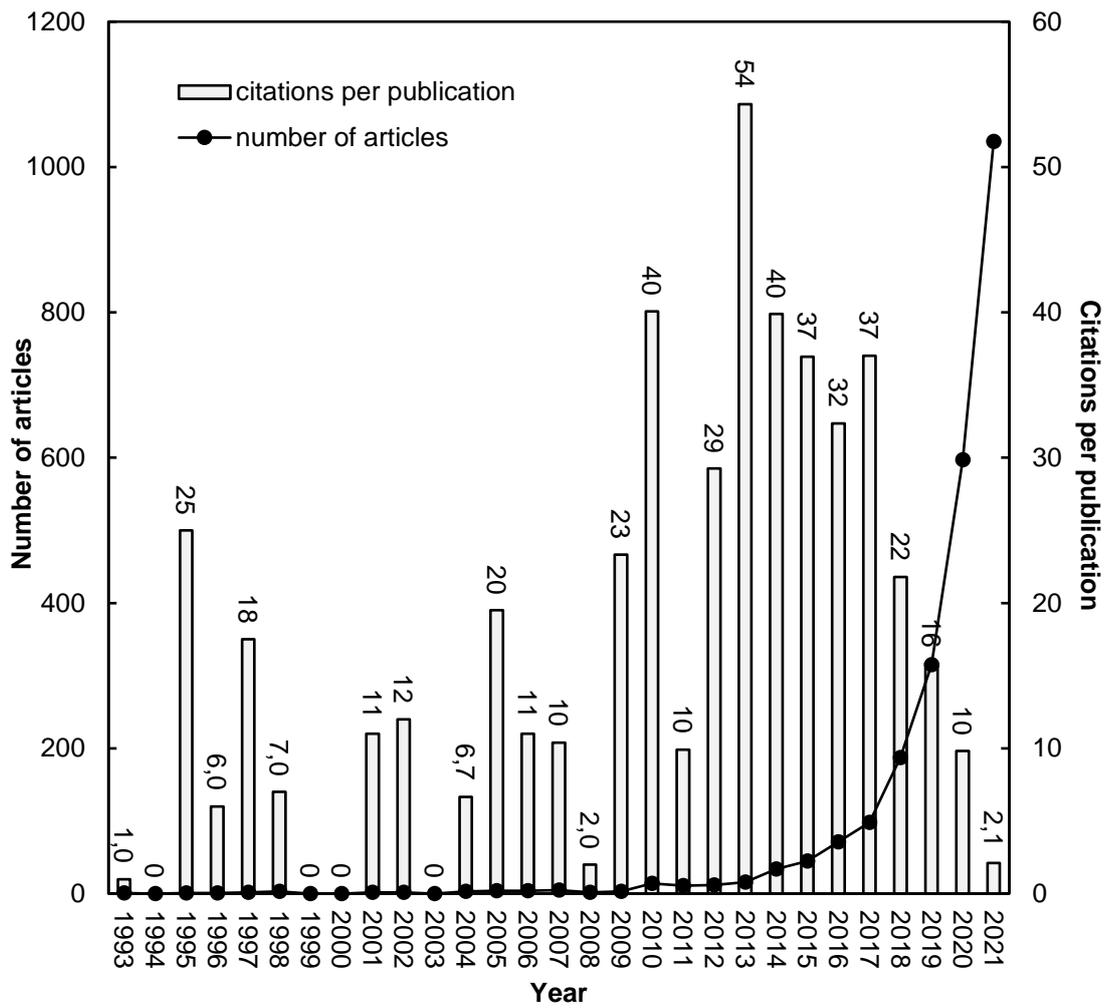

Figure 4: Number of articles and the average number of citations per publication by year.

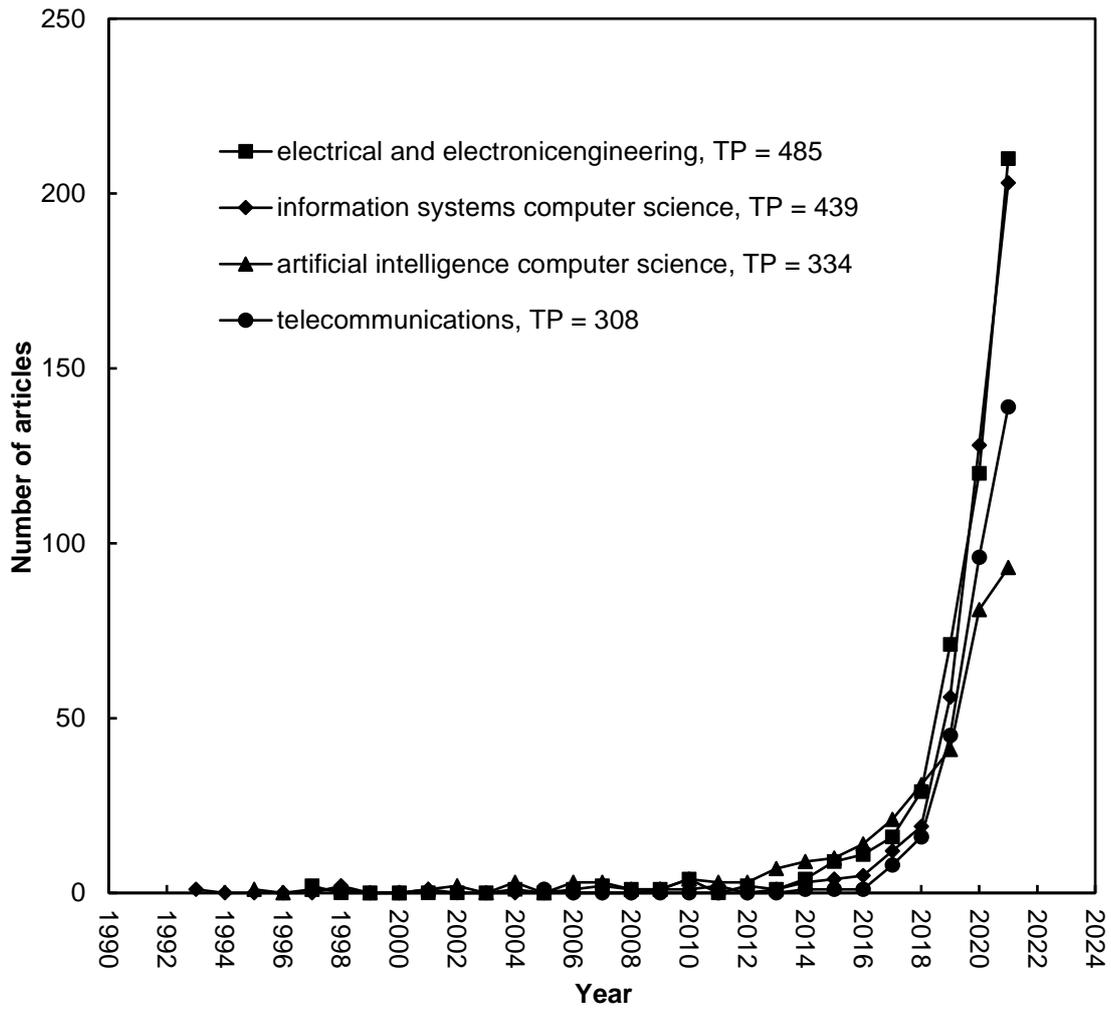

Figure 5. Development of the top four productive Web of Science categories, $TP > 300$.

Table 3. Top 12 most productive journals with $TP > 20$

| Journal | $TP$ (%) | $IF_{2021}$ | APP | $CPP_{2021}$ | Web of Science category |
|---|---|---|---|---|---|
| IEEE Access | 192 (7.8) | 3.476 | 4.5 | 11 | information systems computer science |
| | | | | | electrical and electronic engineering |
| | | | | | telecommunications |
| Remote Sensing | 59 (2.4) | 5.349 | 5.6 | 8.6 | environmental sciences |
| | | | | | multidisciplinary geosciences |
| | | | | | remote sensing |
| | | | | | imaging science and photographic technology |
| PLoS One | 46 (1.9) | 3.752 | 5.7 | 18 | multidisciplinary sciences |
| Sensors | 45 (1.8) | 3.847 | 5.3 | 11 | analytical chemistry |
| | | | | | electrical and electronic engineering |
| | | | | | instruments and instrumentation |
| Scientific Reports | 38 (1.5) | 4.996 | 7.6 | 12 | multidisciplinary sciences |
| Expert Systems With Applications | 37 (1.5) | 8.665 | 3.9 | 30 | artificial intelligence computer science |

| Journal | TP (%) | IF$_{2021}$ | APP | CPP$_{2021}$ | Web of Science Category |
|---|---|---|---|---|---|
| | | | | | electrical and electronic engineering |
| | | | | | operations research and management science |
| Applied Sciences-Basel | 34 (1.4) | 2.838 | 4.7 | 3.2 | multidisciplinary chemistry |
| | | | | | multidisciplinary engineering |
| | | | | | multidisciplinary materials science |
| | | | | | applied physics |
| Monthly Notices of the Royal Astronomical Society | 24 (1.0) | 5.235 | 16 | 18 | astronomy and astrophysics |
| CMC-Computers Materials & Continua | 22 (0.89) | 3.860 | 4.8 | 2.2 | information systems computer science |
| | | | | | multidisciplinary materials science |
| Journal of Big Data | 22 (0.89) | 10.835 | 2.8 | 5.0 | theory and methods computer science |
| Sustainability | 22 (0.89) | 3.889 | 5.1 | 3.5 | green and sustainable science and technology |
| | | | | | environmental sciences |
| | | | | | environmental studies |
| Energies | 21 (0.85) | 3.252 | 4.4 | 6.7 | energy and fuels |

*TP*: total number of articles; %: percentage of articles in all articles; *IF*$_{2021}$: journal impact factor in 2021; *APP*: average number of authors per publication; *CPP*$_{2021}$: average number of citations per paper ($TC_{2021}/TP$).

## 3.4. Publication performances: Countries

Altogether, 649 articles (26% of 2,468 articles) were single-country articles from 16 African countries with an $IP_C$-$CPP_{2021}$ of 10 and 1,819 (74%) were internationally collaborative articles from 146 countries, including 43 African countries and 103 non-African countries with a $CP_C$-$CPP_{2021}$ of 12. The results show citations by international collaborations increased slightly. Six publication indicators and six related citation indicators ($CPP_{2021}$) (Ho and Mukul, 2021) were applied to compare the 44 African countries (Table 4). Egypt dominated in all the six publication indicators with a $TP$ of 777 articles (31% of 2,468 articles), an $IP_C$ of 186 articles (29% of 649 single-country articles), a $CP_C$ of 591 articles (32% of 1,819 internationally collaborative articles), an $FP$ of 345 articles (14% of 2,468 first-author articles), an $RP$ of 449 articles (18% of 2,467 corresponding-author articles), and an $SP$ of 21 articles (32% of 66 single-author articles). Compared to the top 17 productive countries with 20 articles or more, Sudan had a $TP$ of 33 articles, an $IP$ of 3 articles, a $CP$ of 30 articles, an $FP$ of 6 articles, and an $SP$ of 3 articles, with the greatest $TP$-$CPP_{2021}$ of 20, $IP_C$-$CPP_{2021}$ of 23, $CP_C$-$CPP_{2021}$ of 20, $FP$-$CPP_{2021}$ of 14, and $SP$-$CPP_{2021}$ of 23 respectively. Libya had an $FP$ of 3 articles and an $RP$ of 4, with the greatest $FP$-$CPP_{2021}$ of 14 and $RP$-$CPP_{2021}$ of 23. Ten of the 54 African countries such as Angola, Cape Verde, Central African Republic (Cent Afr Republ), Comoros, Djibouti, Equatorial Guinea (Equat Guinea), Eritrea, Sao Tome and Principe (Sao Tome & Prin), Seychelles, and South Sudan had no machine learning-related articles in SCI-EXPANDED. Among the 44 African countries that published machine learning-related articles, 28 countries (64% of 44 African countries) had no single-country articles, while only Niger had no internationally collaborative articles. Similarly, 14 (32%), 10 (23%), and 35 (80%) countries had no first-author, corresponding-author, and single-author articles, respectively.

Development trends in the publication of the top six productive countries with more than 100 articles are presented in Figure 6. From the results obtained, the first machine learning-related article in Africa

(by Egypt) dates back to 1993. In 1995, 1998, 2001, 2004, and 2009, the first articles were published by South Africa, Tunisia, Morocco, Algeria, and Nigeria, respectively. Egypt and South Africa had similar development trends. However, Egypt sharply increased in the last three years to reach 324 articles in 2021. Algeria and Tunisia also had similar development trends.

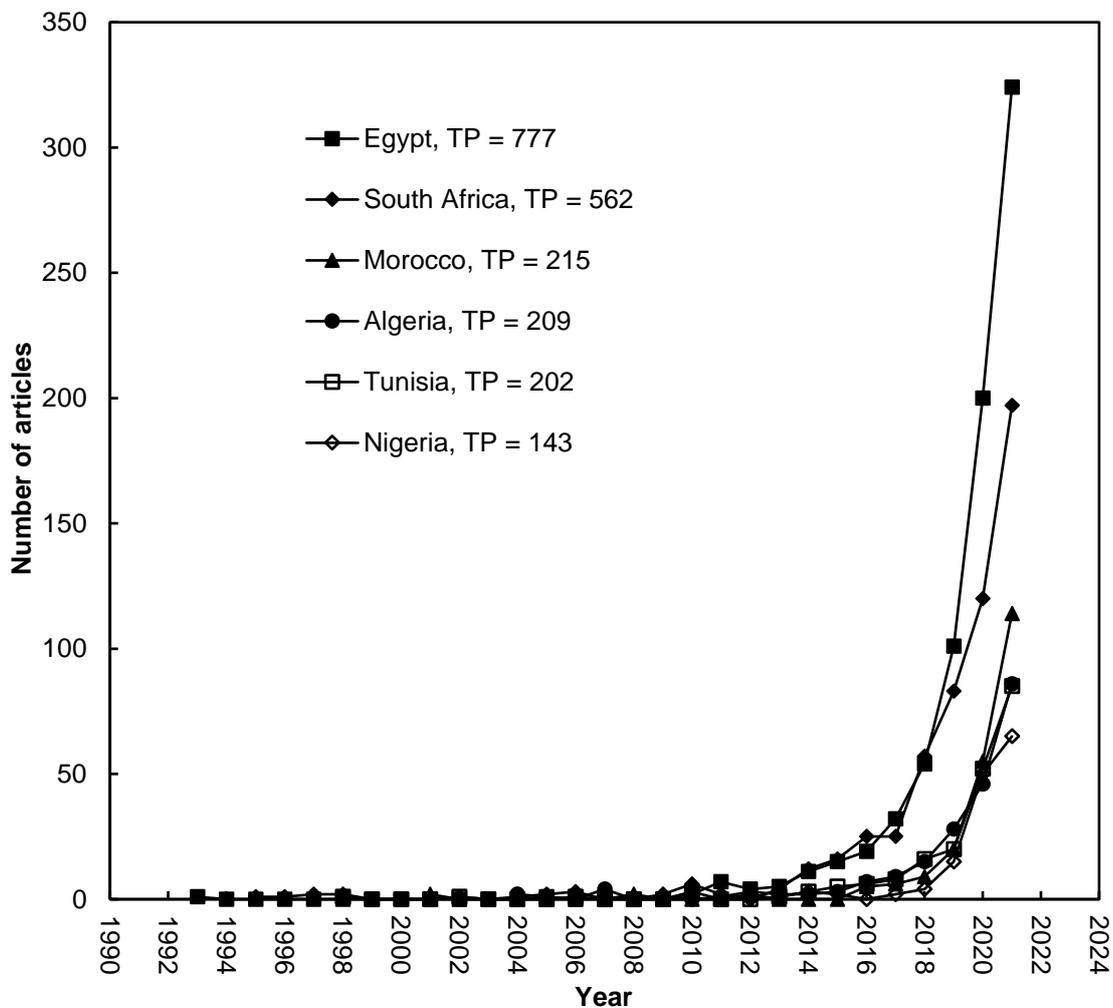

Figure 6. Development of the top six productive countries with $TP > 100$.

Ten of the 103 non-African countries had 100 internationally collaborative articles or more with Africa, as shown in Figure 7. The USA had a $CP_C$ of 431 articles with $CP_C$-$CPP_{2021}$ of 15, followed by Saudi Arabia ($CP_C$ of 338 articles; $CP_C$-$CPP_{2021}$ of 9.3), the UK (295 articles; 14), China (252; 14), France (211; 10), India (174; 11), Germany (156; 20), Canada (154; 14), Australia (146; 13), and Spain (124; 14).

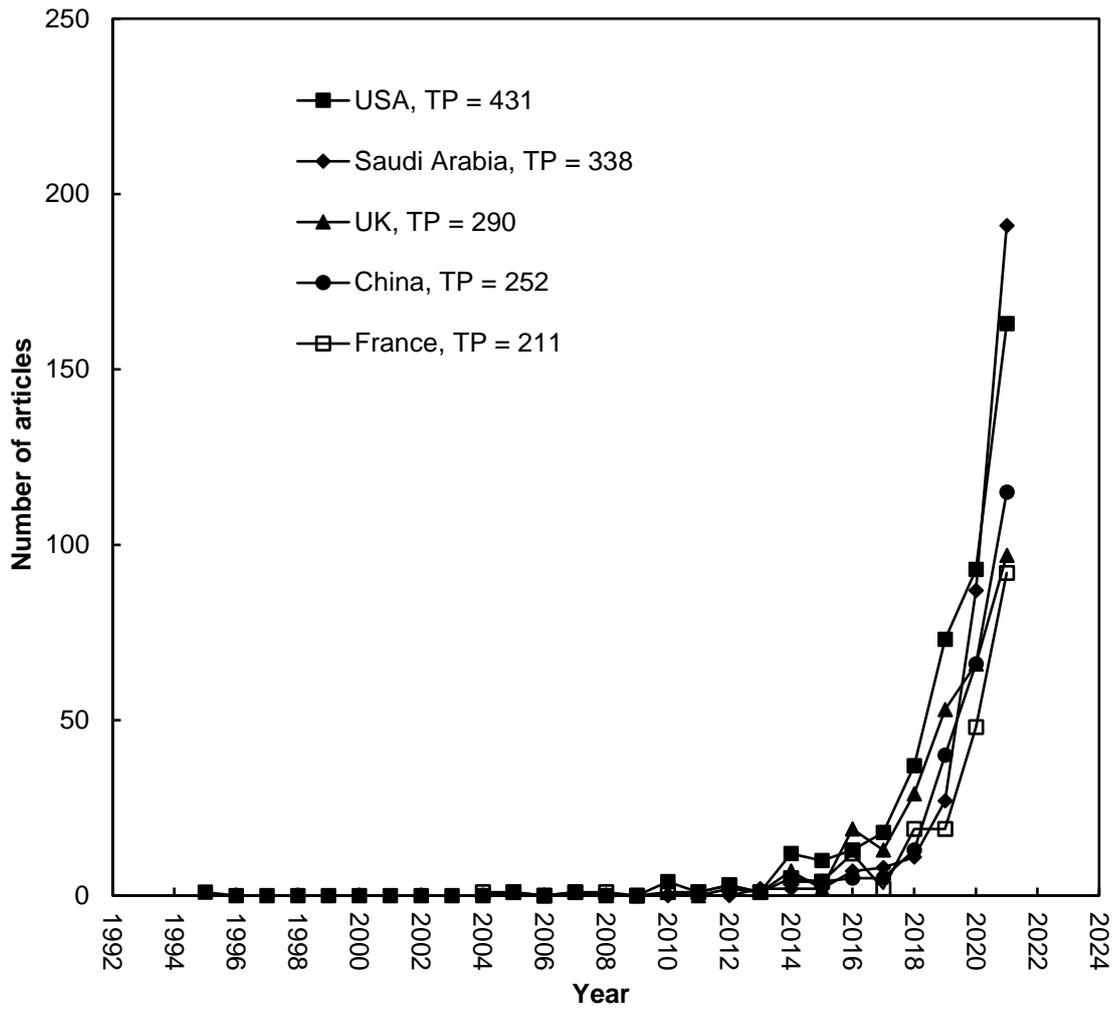

Figure 7. Development of the top five most collaborative countries with Africa, $TP > 200$.

Table 4. African countries published machine learning articles

| Country | TP | TP | | IP$_C$ | | CP$_C$ | | FP | | RP | | SP | |
|---|---|---|---|---|---|---|---|---|---|---|---|---|---|
| | | *TPR* (%) | CPP$_{21}$ | *IP$_C$R* (%) | CPP$_{21}$ | *CP$_C$R* (%) | CPP$_{21}$ | *FPR* (%) | CPP$_{21}$ | *RPR* (%) | CPP$_{21}$ | *SPR* (%) | CPP$_{21}$ |
| Egypt | 777 | 1 (31) | 13 | 1 (29) | 10 | 1 (32) | 14 | 1 (14) | 11 | 1 (18) | 11 | 1 (32) | 8.6 |
| South Africa | 562 | 2 (23) | 14 | 2 (28) | 12 | 2 (21) | 14 | 2 (11) | 12 | 2 (13) | 14 | 2 (21) | 18 |
| Morocco | 215 | 3 (8.7) | 9.1 | 3 (15) | 13 | 6 (6.4) | 5.8 | 4 (6.2) | 10 | 3 (5.9) | 10 | 8 (3.0) | 2.0 |
| Algeria | 209 | 4 (8.5) | 12 | 5 (10) | 9.2 | 3 (8.0) | 13 | 3 (6.5) | 10 | 4 (5.3) | 8.6 | 3 (17) | 5.5 |
| Tunisia | 202 | 5 (8.2) | 7.6 | 4 (10) | 7.8 | 4 (7.5) | 7.5 | 5 (5.1) | 8.4 | 5 (4.9) | 7.1 | 5 (6.1) | 4.0 |
| Nigeria | 143 | 6 (5.8) | 9.0 | 6 (2.8) | 3.3 | 5 (6.9) | 10 | 6 (2.1) | 5.9 | 6 (2.6) | 5.8 | 4 (9.1) | 1.3 |
| Ethiopia | 86 | 7 (3.5) | 5.5 | 8 (1.2) | 3.8 | 7 (4.3) | 5.7 | 8 (0.93) | 6.1 | 7 (2.0) | 4.0 | 6 (4.5) | 2.0 |
| Kenya | 78 | 8 (3.2) | 13 | 11 (0.46) | 0.67 | 8 (4.1) | 14 | 9 (0.85) | 5.8 | 9 (0.89) | 5.3 | N/A | N/A |
| Ghana | 63 | 9 (2.6) | 8.8 | 7 (1.7) | 4.2 | 9 (2.9) | 10 | 7 (1.1) | 4.1 | 8 (1.3) | 7.8 | 8 (3) | 0 |
| Tanzania | 44 | 10 (1.8) | 12 | N/A | N/A | 10 (2.4) | 12 | 11 (0.28) | 4.9 | 12 (0.36) | 4.4 | N/A | N/A |
| Sudan | 33 | 11 (1.3) | 20 | 11 (0.46) | 23 | 11 (1.6) | 20 | 12 (0.24) | 14 | 11 (0.45) | 16 | 6 (4.5) | 23 |
| Uganda | 33 | 11 (1.3) | 5.3 | 9 (0.92) | 10 | 12 (1.5) | 4.3 | 10 (0.45) | 8.3 | 10 (0.53) | 7.2 | N/A | N/A |

| Country | | | | | | | | | | | | | |
|---|---|---|---|---|---|---|---|---|---|---|---|---|---|
| Cameroon | 21 | 13 (0.85) | 10 | N/A | N/A | 13 (1.2) | 10 | 14 (0.2) | 7.4 | 13 (0.32) | 7.1 | N/A | N/A |
| Libya | 21 | 13 (0.85) | 11 | N/A | N/A | 13 (1.2) | 11 | 19 (0.12) | 14 | 18 (0.16) | 23 | N/A | N/A |
| Rwanda | 21 | 13 (0.85) | 4.2 | N/A | N/A | 13 (1.2) | 4.2 | 12 (0.24) | 3.5 | 13 (0.32) | 3.0 | N/A | N/A |
| Zambia | 21 | 13 (0.85) | 6.3 | N/A | N/A | 13 (1.2) | 6.3 | 16 (0.16) | 3.3 | 16 (0.20) | 4.4 | N/A | N/A |
| Zimbabwe | 20 | 17 (0.81) | 7.1 | 14 (0.15) | 5.0 | 17 (1) | 7.2 | 16 (0.16) | 4.8 | 13 (0.32) | 12 | N/A | N/A |
| Botswana | 19 | 18 (0.77) | 15 | 13 (0.31) | 6.0 | 18 (0.93) | 16 | 21 (0.081) | 6.0 | 21 (0.081) | 6.0 | N/A | N/A |
| Senegal | 15 | 19 (0.61) | 10 | N/A | N/A | 19 (0.82) | 10 | 16 (0.16) | 12 | 21 (0.081) | 15 | N/A | N/A |
| Cote Ivoire | 13 | 20 (0.53) | 6.4 | N/A | N/A | 20 (0.71) | 6.4 | 21 (0.081) | 25 | 21 (0.081) | 25 | N/A | N/A |
| Burkina Faso | 11 | 21 (0.45) | 39 | 14 (0.15) | 114 | 22 (0.55) | 32 | 19 (0.12) | 91 | 19 (0.12) | 38 | N/A | N/A |
| Dem Rep Congo | 11 | 21 (0.45) | 4.1 | N/A | N/A | 21 (0.60) | 4.1 | N/A | N/A | N/A | N/A | N/A | N/A |
| Malawi | 10 | 23 (0.41) | 7.9 | N/A | N/A | 22 (0.55) | 7.9 | N/A | N/A | 21 (0.081) | 23 | N/A | N/A |
| Mozambique | 10 | 23 (0.41) | 8.0 | N/A | N/A | 22 (0.55) | 8.0 | 25 (0.041) | 14 | 21 (0.081) | 24 | N/A | N/A |
| Madagascar | 9 | 25 (0.36) | 21 | N/A | N/A | 25 (0.49) | 21 | 21 (0.081) | 3.5 | 19 (0.12) | 2.7 | N/A | N/A |
| Mauritius | 8 | 26 (0.32) | 5.0 | 10 (0.62) | 2.0 | 29 (0.22) | 8.0 | 14 (0.2) | 2.0 | 16 (0.2) | 2.0 | N/A | N/A |
| Benin | 6 | 27 (0.24) | 4.7 | N/A | N/A | 26 (0.33) | 4.7 | 21 (0.081) | 2.5 | 21 (0.081) | 2.5 | N/A | N/A |
| Gambia | 6 | 27 (0.24) | 14 | N/A | N/A | 26 (0.33) | 14 | N/A | N/A | 28 (0.041) | 2.0 | N/A | N/A |

| Country | | | | | | | | | | | | |
|---|---|---|---|---|---|---|---|---|---|---|---|---|
| Sierra Leone | 6 | 27 (0.24) | 9.3 | N/A | N/A | 26 (0.33) | 9.3 | 25 (0.041) | 3.0 | 21 (0.081) | 3.0 | N/A | N/A |
| Gabon | 4 | 30 (0.16) | 12 | N/A | N/A | 29 (0.22) | 12 | N/A | N/A | 28 (0.041) | 5.0 | N/A | N/A |
| Mali | 4 | 30 (0.16) | 30 | N/A | N/A | 29 (0.22) | 30 | N/A | N/A | 28 (0.041) | 8.0 | N/A | N/A |
| Namibia | 4 | 30 (0.16) | 21 | N/A | N/A | 29 (0.22) | 21 | N/A | N/A | N/A | N/A | N/A | N/A |
| Guinea | 3 | 33 (0.12) | 5.0 | N/A | N/A | 33 (0.16) | 5.0 | 25 (0.041) | 4.0 | 28 (0.041) | 4.0 | N/A | N/A |
| Togo | 3 | 33 (0.12) | 5.0 | N/A | N/A | 33 (0.16) | 5.0 | 25 (0.041) | 11 | 28 (0.041) | 11 | N/A | N/A |
| Burundi | 2 | 35 (0.081) | 1.5 | N/A | N/A | 35 (0.11) | 1.5 | N/A | N/A | N/A | N/A | N/A | N/A |
| Chad | 2 | 35 (0.081) | 4.5 | N/A | N/A | 35 (0.11) | 4.5 | N/A | N/A | N/A | N/A | N/A | N/A |
| Somalia | 2 | 35 (0.081) | 10 | N/A | N/A | 35 (0.11) | 10 | 25 (0.041) | 8.0 | 28 (0.041) | 11 | N/A | N/A |
| Eswatini | 1 | 38 (0.041) | 0 | N/A | N/A | 38 (0.055) | 0 | N/A | N/A | N/A | N/A | N/A | N/A |
| Guinea Bissau | 1 | 38 (0.041) | 28 | N/A | N/A | 38 (0.055) | 28 | N/A | N/A | N/A | N/A | N/A | N/A |
| Lesotho | 1 | 38 (0.041) | 5.0 | N/A | N/A | 38 (0.055) | 5.0 | N/A | N/A | N/A | N/A | N/A | N/A |
| Liberia | 1 | 38 (0.041) | 5.0 | N/A | N/A | 38 (0.055) | 5.0 | N/A | N/A | N/A | N/A | N/A | N/A |
| Mauritania | 1 | 38 (0.041) | 1.0 | N/A | N/A | 38 (0.055) | 1.0 | N/A | N/A | N/A | N/A | N/A | N/A |
| Niger | 1 | 38 (0.041) | 1.0 | 14 (0.15) | 1.0 | N/A | N/A | 25 (0.041) | 1.0 | 28 (0.041) | 1.0 | N/A | N/A |
| Rep Congo | 1 | 38 (0.041) | 18 | N/A | N/A | 38 (0.055) | 18 | N/A | N/A | N/A | N/A | N/A | N/A |

*TP*: total number of articles; *TPR* (%): rank of the total number of articles and percentage in 2,468 articles; *IP*$_C$*R* (%): rank of single-country articles and percentage in 649 single-country articles; *CP*$_C$*R* (%): rank of internationally collaborative articles and percentage in 1,819 internationally collaborative articles; *FPR* (%): rank of first-author articles and percentage in 2,468 first-author articles; *RPR* (%): rank of corresponding-author articles and percentage in 2,467 corresponding-author articles; *SPR* (%): rank of single-author articles and percentage in 66 single-author articles; *CPP*$_{21}$: average number of citations per publication (*TC*$_{2021}$/*TP*); N/A: not available.

## 3.5. Publication performances: Institutions

Concerning institutions, 382 African articles (15% of 2,468 articles) originated from single institutions with an $IP_I$-$CPP_{2021}$ of 9.9, while 2,086 articles (85%) were institutional collaborations with a $CP_I$-$CPP_{2021}$ of 12. The institutional collaborations slightly increased the citations. The top 20 productive African institutions and their characteristics are presented in Table 5. Cairo University in Egypt ranked top with a $TP$ of 142 articles (5.8% of 2,468 articles) and a $CP_I$ of 127 articles (6.1% of 2,086 inter-institutionally collaborative articles). However, the University of KwaZulu-Natal in South Africa ranked top in three of the six publication indicators with an $IP$ of 19 articles (5.0% of 382 single-institution articles), an $FP$ of 48 articles (1.9% of 2,468 first-author articles), and an $RP$ of 64 articles (2.6% of 2,467 corresponding-author articles). In addition, the University of Johannesburg in South Africa and the Council of Scientific and Industrial Research (CSIR) in South Africa ranked top with an $SP$ of four articles (6.1% of 66 single-author articles), respectively. Compared to the top 20 African countries, the University of KwaZulu-Natal in South Africa had a $TP$ of 104 articles, a $CP_I$ of 85 articles, an $FP$ of 48 articles, and an $RP$ of 64 articles, with the greatest $TP$-$CPP_{2021}$ of 24, $CP_I$-$CPP_{2021}$ of 27, $FP$-$CPP_{2021}$ of 27, and $RP$-$CPP_{2021}$ of 34 respectively. The University of Pretoria in South Africa had an $IP_I$ of 15 articles with the greatest $IP_I$-$CPP_{2021}$ of 20, while the Mansoura University in Egypt had an $SP$ of two articles with the greatest $SP$-$CPP_{2021}$ of 29.

Five non-Africa institutions had 30 inter-institutionally collaborative articles or more with Africa. King Saud University in Saudi Arabia had a $CP_I$ of 62 articles with $CP_I$-$CPP_{2021}$ of 10, followed by Taif University in Saudi Arabia ($CP_I$ of 39 articles; $CP_I$-$CPP_{2021}$ of 2.3), University of Oxford in the UK (38 articles; 15), King Abdulaziz University in Saudi Arabia (34; 4.1), and Prince Sattam Bin Abdulaziz University in Saudi Arabia (31; 7.6).

Table 5. Top 20 most productive African institutions

| Institution | TP | TP | | IP$_I$ | | CP$_I$ | | FP | | RP | | SP | |
|---|---|---|---|---|---|---|---|---|---|---|---|---|---|
| | | R (%) | CPP | R (%) | CPP | R (%) | CPP | R (%) | CPP | R (%) | CPP | R (%) | CPP |
| CU, Egypt | 142 | 1 (5.8) | 17 | 3 (3.9) | 18 | 1 (6.1) | 17 | 4 (1.3) | 23 | 2 (2.1) | 14 | 5 (3.0) | 7.0 |
| UKZN, South Africa | 104 | 2 (4.2) | 24 | 1 (5.0) | 8.8 | 3 (4.1) | 27 | 1 (1.9) | 27 | 1 (2.6) | 34 | 3 (4.5) | 5.0 |
| UCT, South Africa | 101 | 3 (4.1) | 14 | 10 (1.8) | 5.9 | 2 (4.5) | 15 | 6 (0.93) | 10 | 4 (1.3) | 6.6 | 12 (1.5) | 5.0 |
| MansU, Egypt | 92 | 4 (3.7) | 12 | 5 (3.4) | 10 | 4 (3.8) | 12 | 2 (1.5) | 8.1 | 3 (1.9) | 10 | 5 (3.0) | 29 |
| ZU, Egypt | 69 | 5 (2.8) | 16 | 27 (0.79) | 5.0 | 5 (3.2) | 17 | 8 (0.81) | 12 | 7 (1.1) | 10 | N/A | N/A |
| UW, South Africa | 66 | 6 (2.7) | 11 | 19 (1.0) | 10 | 6 (3.0) | 11 | 8 (0.81) | 4.8 | 12 (0.89) | 5.5 | N/A | N/A |
| BU, Egypt | 64 | 7 (2.6) | 13 | 36 (0.52) | 0.50 | 6 (3.0) | 13 | 14 (0.57) | 24 | 9 (1.0) | 18 | 12 (1.5) | 1.0 |
| MU, Egypt | 60 | 8 (2.4) | 10 | 27 (0.79) | 1.3 | 8 (2.7) | 11 | 20 (0.41) | 2.8 | 14 (0.81) | 10 | N/A | N/A |
| UP, South Africa | 59 | 9 (2.4) | 12 | 3 (3.9) | 20 | 10 (2.1) | 9.4 | 5 (1.3) | 13 | 4 (1.3) | 12 | N/A | N/A |
| ASU, Egypt | 55 | 10 (2.2) | 14 | 9 (2.1) | 3.3 | 9 (2.3) | 15 | 7 (0.89) | 19 | 7 (1.1) | 16 | 5 (3.0) | 2.0 |
| UJ, South Africa | 54 | 11 (2.2) | 5.1 | 2 (4.7) | 8.1 | 12 (1.7) | 3.6 | 3 (1.4) | 7.1 | 6 (1.3) | 7.4 | 1 (6.1) | 7.8 |
| UWC, South Africa | 48 | 12 (1.9) | 11 | 14 (1.3) | 13 | 11 (2.1) | 11 | 17 (0.53) | 7.8 | 15 (0.69) | 8.9 | 12 (1.5) | 20 |

| Institution | TP | TP R (%) | IPI R (%) | CPI R (%) | FP R (%) | RP R (%) | SPI R (%) | TP-CPP | IPI-CPP | CPI-CPP | FP-CPP | RP-CPP |
|---|---|---|---|---|---|---|---|---|---|---|---|---|
| SU, South Africa | 42 | 13 (1.7) | 7.8 | 8 (2.4) | 7.9 | 13 (1.6) | 7.7 | 12 (0.77) | 9.2 | 9 (1) | 7.2 | 12 (1.5) | 0 |
| HU, Egypt | 36 | 14 (1.5) | 10 | 19 (1.0) | 19 | 14 (1.5) | 8.6 | 18 (0.49) | 15 | 19 (0.61) | 12 | 12 (1.5) | 1.0 |
| TU, Egypt | 33 | 15 (1.3) | 12 | 66 (0.26) | 9.0 | 14 (1.5) | 12 | 36 (0.28) | 6.0 | 27 (0.45) | 11 | N/A | N/A |
| UTEM, Tunisia | 33 | 15 (1.3) | 3.1 | 14 (1.3) | 1.2 | 16 (1.3) | 3.5 | 8 (0.81) | 3.8 | 11 (1.0) | 1.7 | N/A | N/A |
| AU, Egypt | 30 | 17 (1.2) | 9.0 | 36 (0.52) | 10 | 16 (1.3) | 8.9 | 55 (0.16) | 7.3 | 24 (0.53) | 6.9 | 12 (1.5) | 7.0 |
| UT, Tunisia | 28 | 18 (1.1) | 10 | 13 (1.6) | 5.3 | 23 (1.1) | 11 | 8 (0.81) | 11 | 13 (0.85) | 11 | 12 (1.5) | 5.0 |
| SCU, Egypt | 27 | 19 (1.1) | 23 | 36 (0.52) | 3.0 | 19 (1.2) | 24 | 37 (0.24) | 10 | 19 (0.61) | 21 | N/A | N/A |
| UC, Tunisia | 27 | 19 (1.1) | 8.6 | 10 (1.8) | 6.3 | 26 (1.0) | 9.4 | 23 (0.36) | 5.1 | 22 (0.57) | 14 | 3 (4.5) | 3.7 |

*TP*: total number of articles; *TP R* (%): total number of articles and percentage of total articles; *IP$_I$ R* (%): rank and percentage of single-institution articles in all single-institution articles; *CP$_I$ R* (%): rank and percentage of inter-institutionally collaborative articles in all inter-institutionally collaborative articles; *FP R* (%): rank and percentage of first-author articles in all first-author articles; *RP R* (%): rank and percentage of corresponding-author articles in all corresponding-author articles; *SP$_I$ R* (%): rank and percentage of single-author articles in all single-author articles; *TP-CPP*: the total *TC*$_{2021}$ of all articles per the total number of articles (*TP*); *IP$_I$-CPP*: the total *TC*$_{2021}$ of all single-institution articles per the number of single-institution articles (*IP$_I$*); *CP$_I$-CPP*: the total *TC*$_{2021}$ of all inter-institutionally collaborative articles per the number of inter-institutionally collaborative articles (*CP$_I$*); *FP-CPP*: the total *TC*$_{2021}$ of all first-author per the number of first-author articles (*FP*); *RP-CPP*: the total *TC*$_{2021}$ of all corresponding-author articles per the number of corresponding-author articles (*RP*); N/A: not available.

## 3.6. Citation histories of the ten most frequently cited articles

The total citations in the Web of Science Core Collection are updated from time to time. To improve bibliometric studies directly using data from the database, total citations from the Web of Science Core Collection from the year of publication to the end of the most recent year of 2021 ($TC_{2021}$) were applied (Wang et al., 201; Al-Moraissi et al., 2022). The citation history of the most frequently cited articles assessed by $TC_{year}$ in a research topic was presented to understand the impact history of the articles (Ho, 2012; Al-Moraissi et al., 2022; Mohsen & Ho et al., 2022). Highly cited articles may not always significantly impact a research field (Ho, 2014a; b; Mohsen & Ho et al., 2022). Table 6 shows the top ten most frequently cited machine learning-related articles in Africa. Five of the top ten articles were published by Egypt, followed by South Africa with two articles and one each by Nigeria, Kenya, and Morocco.

The most cited article was entitled *Ranger: a fast implementation of random forests for high dimensional data in C++ and R* (Wright and Ziegler, 2017) by M.N. Wright and A. Ziegler from the University of Lubeck in Germany and the University of KwaZulu-Natal in South Africa and had a $TC_{2021}$ of 683 (rank 1st) and a $C_{2021}$ of 330 (rank 2nd). An article entitled *Peeking inside the black-box: A survey on explainable artificial intelligence (XAI)* (Adadi and Berrada, 2018) by A. Adadi and M. Berrada from the Sidi Mohammed Ben Abdellah University in Morocco had the most impact on the most recent year of 2021 with a $C_{2021}$ of 435 (rank 1st) and a $TC_{2021}$ of 675 (rank 2nd). These two articles keep increasing in citations.

Table 6. The top ten most frequently cited articles by African countries

| Rank ($TC_{2021}$) | Rank ($C_{2021}$) | Title | Country | Reference |
|---|---|---|---|---|
| 1 (683) | 2 (330) | Ranger: A fast implementation of random forests for high dimensional data in C++ and R | Germany, South Africa | Wright and Ziegler (2017) |
| 2 (675) | 1 (435) | Peeking inside the black-box: A survey on explainable artificial intelligence (XAI) | Morocco | Adadi and Berrada (2018) |
| 3 (402) | 22 (56) | Multiobjective intelligent energy management for a microgrid | Japan, Egypt, Saudi Arabia | Chaouachi et al. (2013) |
| 4 (313) | 31 (47) | Computer-aided diagnosis of human brain tumor through MRI: A survey and a new algorithm | Egypt, UK | El-Dahshan et al. (2014) |
| 5 (255) | 8 (87) | Predicting carbon dioxide and energy fluxes across global FLUXNET sites with regression algorithms | Italy, Germany, USA, Japan, Spain, Romania, Canada, Ireland, Switzerland, Kenya | Tramontana et al. (2016) |
| 6 (251) | 15 (64) | An introduction to quantum machine learning | South Africa | Schuld et al. (2015) |
| 7 (228) | 20 (59) | An empirical comparison of machine learning models for time series forecasting | Egypt, USA | Ahmed et al. (2010) |
| 8 (200) | 48 (35) | A support vector machine: Firefly algorithm-based model for global solar radiation prediction | Malaysia, Nigeria, Iran, Serbia, India | Olatomiwa et al. (2015) |
| 9 (199) | 13 (73) | Machine learning with big data: Challenges and approaches | Canada, Egypt | L'Heureux et al. (2017) |
| 10 (185) | 6 (94) | Linear discriminant analysis: A detailed tutorial | Germany, Egypt | Tharwat et al. (2017) |

$TC_{2021}$: number of citations from Web of Science Core Collection since its publication to the end of 2021; $C_{2021}$: number of citations from Web of Science Core Collection in 2021.

## 3.7. Research foci

In the last decade, Ho's research group proposed distributions of words in article titles and abstracts, author keywords, and *Keywords Plus* of different periods to determine research foci and trends (Mao et al., 2010; Wang and Ho, 2016). Among 2,468 articles, 2,464 articles (99.8% of 2,468 articles) had record information of article abstracts; 2,103 (85.2%) articles had author keywords; and 2,069 (83.8%) articles had *Keywords Plus*. The 20 most frequent keywords are listed in Table 7. The classification was ranked in the top 20 in article titles and abstracts, author keywords, and *Keywords Plus*, respectively. The development of the top four topics in machine learning in Africa, such as deep learning, classification, feature extraction, and random forest, is shown in Figure 8.

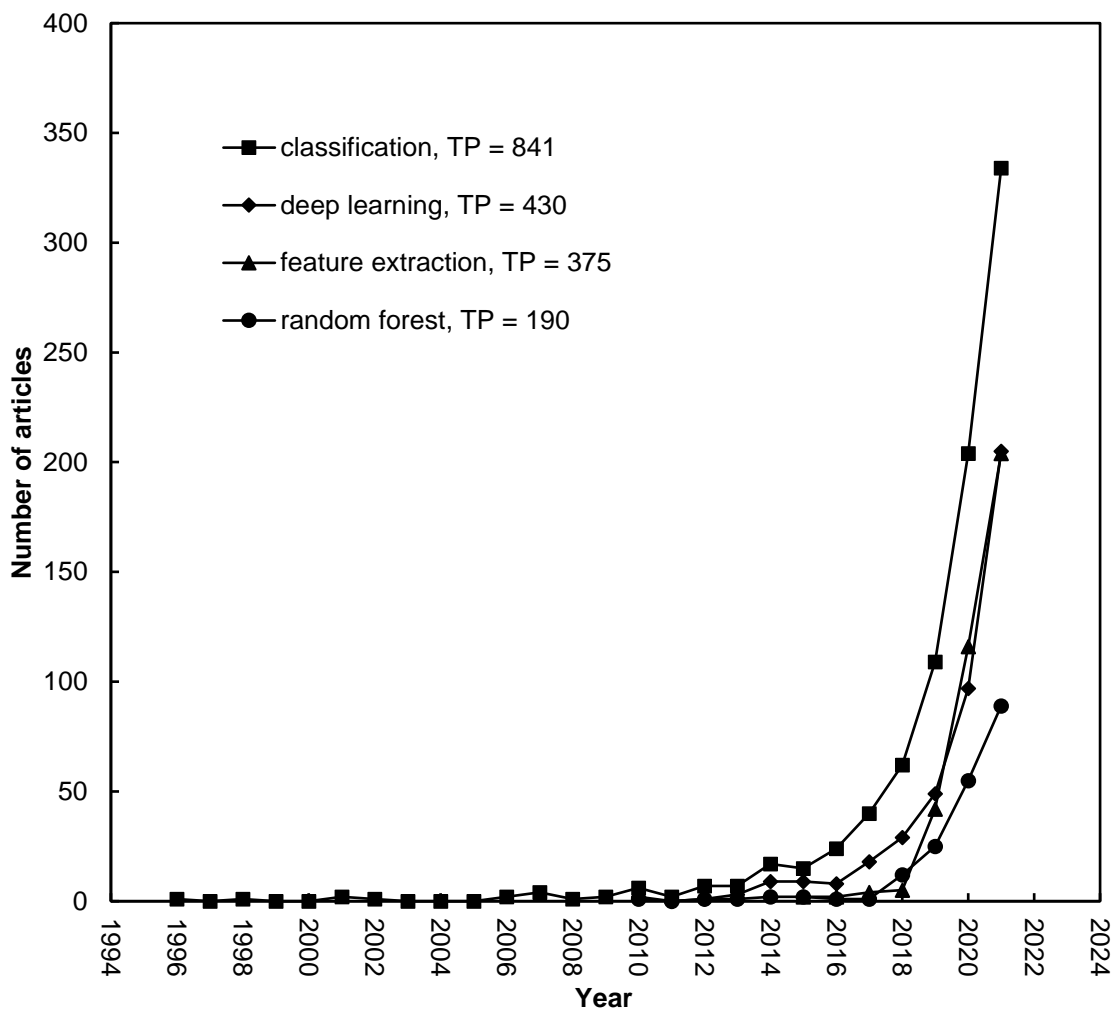

Figure 8. Development trends of the four most popular topics in Africa

*i. Classification*

Articles containing supporting words such as classification, classifications, and misclassification in their title, abstract, or author keywords were classified as classification-related articles. In 1996, F.S. Gouws and Aldrich from the University of Stellenbosch in South Africa reported that using machine learning techniques and the classification rules on a supervisory expert system shell or decision support system for plant operators could consequently make a significant impact on the way notation plants (Gouws and Aldrich, 1996). Highly cited articles with $TC_{2021}$ of 100 or more (Ho, 2014b), such as *Deep learning for tomato diseases: Classification and symptoms visualization* (Brahimi et al., 2017) and *Learning machines and sleeping brains: Automatic sleep stage classification using decision-tree multi-class support vector machines* (Lajnef et al., 2015) were published by African authors from Algeria and Tunisia respectively. An article entitled *A predictive machine learning application in agriculture: Cassava disease detection and classification with imbalanced dataset using convolutional neural networks* (Sambasivam and Opiyo, 2021) was published in the most recent year 2021 by Sambasivam and Opiyo from International Business, Science And Technology University (ISBAT) in Uganda.

*ii. Deep learning*

Supporting words for deep learning were deep learning, deep neural network, deep neural networks, deep transfer learning, deep reinforcement learning, deep convolutional neural network, and deep convolutional neural networks. Deep learning was first mentioned in an article on *Deep learning framework with confused sub-set resolution architecture for automatic Arabic Diacritization* (Rashwan et al., 2015) by authors from Egypt and Kuwait. Highly cited machine learning article was published by African authors, for example, *Deep learning for tomato diseases: Classification and symptoms visualization* (Brahimi et al., 2017) by authors from Algeria and *Deep learning for cyber*

*security intrusion detection: Approaches, datasets, and comparative study* (Ferrag et al., 2020) by authors from Algeria and the UK. The most impactful article about deep learning in 2021 was *A hybrid deep transfer learning model with machine learning methods for face mask detection in the era of the COVID-19 pandemic* (Loey et al., 2021) by authors from Egypt, USA, and Taiwan.

*iii. Feature extraction*

Supporting words for the feature extraction were feature extraction, feature selection, and feature evaluation. Saidi et al. from France and Tunisia published the first feature extraction-related article entitled *Protein sequences classification by means of feature extraction with substitution matrices* (Saidi et al., 2010) in Africa. Highly cited articles about feature extraction were *Ensemble-based multi-filter feature selection methods for DDoS detection in cloud computing* (Osanaiye et al., 2016) by authors from South Africa, Australia, China, and the UK and *Minimum redundancy maximum relevance feature selection approach for temporal gene expression data* (Radovic et al., 2017) by authors from the USA, Serbia, and Egypt. In 2021, *Metaheuristic algorithms on feature selection: A survey of one decade of research (2009-2019)* (Agrawal et al., 2021) was published by authors from India, Saudi Arabia, and Egypt.

*iv. Random forest*

Supporting words for the random forest were random forest, random forests, and random decision forest. In 2010, Auret and Aldrich (2010) from the University of Stellenbosch in South Africa published the first article about the random forest in machine learning. Highly cited random forest-related articles were published in the last decade in Africa, for example, *Ranger: A fast implementation of random forests for high dimensional data in C++ and R* (Wright and Ziegler, 2017) by Wright and Ziegler from Germany and South Africa and *Random forest regression and spectral band selection for estimating sugarcane leaf nitrogen concentration using EO-1 Hyperion hyperspectral data* (Abdel-Rahman et al., 2013) by authors from South Africa and Sudan. In 2021,

*The application of the random forest classifier to map irrigated areas using Google Earth Engine* (Magidi et al., 2021) was presented by authors from South Africa.

The yearly development trends of the four most popular topics in Africa, shown in Figure 8, illustrated that the classification ($TP$ = 841 articles) was the most concerned with machine learning in Africa. Research about deep learning was more popular than feature extraction. However, they have shown the same development trends in recent years.

Table 7. The 20 most frequently used keywords

| Words in title | TP | R (%) | Words in Abstract | TP | R (%) | Author keywords | TP | R (%) | Keywords Plus | TP | R (%) |
|---|---|---|---|---|---|---|---|---|---|---|---|
| learning | 886 | 1 (36) | learning | 1,930 | 1 (78) | machine learning | 918 | 1 (44) | classification | 279 | 1 (13) |
| machine | 739 | 2 (30) | machine | 1,925 | 2 (78) | deep learning | 189 | 2 (9.0) | prediction | 207 | 2 (10) |
| detection | 246 | 3 (10) | model | 1,104 | 3 (45) | classification | 101 | 3 (4.8) | model | 164 | 3 (7.9) |
| classification | 235 | 4 (10) | accuracy | 1,067 | 4 (43) | feature extraction | 101 | 3 (4.8) | algorithm | 116 | 4 (5.6) |
| prediction | 232 | 5 (9.4) | proposed | 1,013 | 5 (41) | random forest | 95 | 5 (4.5) | system | 89 | 5 (4.3) |
| approach | 196 | 6 (7.9) | paper | 928 | 6 (38) | feature selection | 80 | 6 (3.8) | diagnosis | 88 | 6 (4.3) |
| deep | 192 | 7 (7.8) | methods | 868 | 7 (35) | artificial intelligence | 75 | 7 (3.6) | selection | 85 | 7 (4.1) |
| model | 166 | 8 (6.7) | models | 865 | 8 (35) | support vector machine | 59 | 8 (2.8) | regression | 82 | 8 (4) |
| analysis | 155 | 9 (6.3) | approach | 812 | 9 (33) | support vector machines | 55 | 9 (2.6) | neural-networks | 80 | 9 (3.9) |
| neural | 143 | 10 (5.8) | analysis | 780 | 10 (32) | optimization | 54 | 10 (2.6) | performance | 79 | 10 (3.8) |
| system | 128 | 11 (5.2) | classification | 745 | 11 (30) | Covid-19 | 53 | 11 (2.5) | neural-network | 75 | 11 (3.6) |
| algorithms | 123 | 12 (5) | algorithm | 731 | 12 (30) | artificial neural network | 52 | 12 (2.5) | identification | 64 | 12 (3.1) |
| network | 122 | 13 (4.9) | techniques | 714 | 13 (29) | data mining | 45 | 13 (2.1) | models | 63 | 13 (3) |
| networks | 117 | 14 (4.7) | algorithms | 705 | 14 (29) | neural networks | 44 | 14 (2.1) | optimization | 61 | 14 (2.9) |
| algorithm | 115 | 15 (4.7) | method | 703 | 15 (29) | prediction | 44 | 14 (2.1) | design | 60 | 15 (2.9) |
| techniques | 115 | 15 (4.7) | compared | 691 | 16 (28) | remote sensing | 44 | 14 (2.1) | feature-selection | 57 | 16 (2.8) |
| feature | 105 | 17 (4.3) | neural | 622 | 17 (25) | convolutional neural network | 41 | 17 (1.9) | systems | 49 | 17 (2.4) |
| hybrid | 103 | 18 (4.2) | network | 621 | 18 (25) | machine learning algorithms | 41 | 17 (1.9) | features | 46 | 18 (2.2) |
| selection | 98 | 19 (4) | features | 616 | 19 (25) | big data | 37 | 19 (1.8) | support vector machine | 46 | 18 (2.2) |
| models | 96 | 20 (3.9) | support | 604 | 20 (25) | internet of things | 37 | 19 (1.8) | random forest | 43 | 20 (2.1) |

*TP*: number of articles; *R*: rank in a period; N/A: not available

## 4. Machine Learning Research

### 4.1. Preliminary overview

Machine learning (ML) is a subfield of artificial intelligence. The central idea is that the machine learns by interacting with the input data and develops a corresponding model capable of classifying a new input or predicting an outcome based on new inputs. The input data is usually divided into two: the training data used to teach the machine and the classification data used for testing the accuracy of the trained model. Different ML algorithms have been used to solve problems such as early disease detection and classification in medicine and agriculture, plants or crops disease detection, data mining, clustering, quantum computing technology, engineering optimization, earth observation, food security, climate change, pollution, and many more.

### 4.2. Research trends in Africa

ML algorithms have found significant application in bioinformatics, especially in genetic testing of microscopic spots stored in DNA microarrays, genomics, and proteomics. Also, the medical or biological fields have been receiving significant attention from ML researchers, particularly in areas of medical engineering, epidemiology, and the study and early detection of genetic diseases and disorders such as Alzheimer's disease, diabetes, cancer, arthritis, high blood pressure, hemochromatosis, cystic fibrosis, Huntington's disease, sickle cell anemia, and Marfan syndrome (Chou, et al., 2017; Nevado-Holgado & Lovestone, 2017; Mulder, et al., 2017). Focusing on diseases prevalent in Africa, machine learning has been used to improve the genetic resistance to malaria, early detection and eradication of diabetes, classification of sickle cell anemia, improvement of genetic resistance to HIV/AIDS, and detection of uterine fibroids in women (Laughlin, Schroeder, & Baird, 2010).

ML has also found tremendous application in the economy and is argued to be the bedrock of the fourth industrial revolution (Schwab, 2016). The developed countries have keyed into this to avoid missing out on the revolution (AI EU, 2021; AI Japan, 2021). Actors in government and private sectors have developed strategies that key into the revolution. Africa is lagging in this regard with little or no efforts towards actualizing the fourth industrial revolution. Some agencies from the West have tried to assist developing countries (Digital Africa, 2021; FAIR Forward, 2021). Countries like Rwanda and others have taken the initiative of developing plans driven by AI to achieve economic sustainability (AI Rwanda, 2021)

Different AI techniques, such as ML and the Internet of Things (IoT), drive the energy sector. Africa is not left behind in this aspect. ML is used in pay-as-you-go energy products to predict demand, score users' activities, and develop models that make products available, affordable and adaptable (Arakpogun, Elsahn, Olan, & Elsahn, 2021). For example, an energy company can use the predictive analysis aspect of ML to make available energy services or products to areas without access to energy products and services (Equatorial Power, 2021; Quartz Africa, 2021).

The agricultural sector offers a fertile ground for ML to display the ability to improve productivity and efficiency all along the value chain. It provides solutions for subsistence and mechanized farmers to improve yield and increase profits through developing models for the detection and precision treatment of pests and diseases, optimal fertilizer application, soil monitoring, and many more. Solutions like Gro intelligence in Kenya deploy AI techniques such as ML to achieve food security (Gro Intelligence, 2021). Climatic conditions for precision agriculture have been achieved through the use of drone technology with the capability of knowing the optimal interventions needed for optimal yield (Gebbers & Adamchuk, 2010)

ML has also been used to develop systems that could identify in real time the appropriate agronomic interventions that should be made using sensor data such as pH level, soil moisture level, temperature, and more. In Kenya and Mozambique, projects like Third Eye drive this process for better yield (Third Eye, 2021; Badiane & Jv, 2019). Western technologies like Farmbeats have been applied in Africa using low-cost, sparsely distributed sensors and aerial imagery to generate precision maps. The system is attached to a smartphone carrying helium balloons, which is a low-cost drone system (Swamy et al., 2019; Vasisht et al., 2017). Intelligent drones with high ML capabilities have been deployed to survey elephants in Burkina Faso, anti-poaching rhinos in South Africa, and analysis of flood risks in Tanzania (Vermeulen, Lejeune, Lisein, Sawadogo, & Bouché, 2013; Soesilo et al., 2016; Mulero-Pázmány, Stolper, Van Essen, Negro, & Sassen, 2014).

In entrepreneurship, ML has been leveraged to deliver innovative research and products. Hepta Analytics developed a product called Najua, which uses ML to present web content in local languages (Najua, 2021). A start-up company in Nigeria developed a mobile app called Ubenwa, which is used to detect early prenatal asphyxia in newborn babies by analyzing acoustic signatures (Onu et al., 2017)

### 4.3. Major application areas

ML is a major driver of the Fourth Industrial Revolution (4IR). It has improved outcomes in various application areas by utilizing its learning and prediction abilities. This section summarizes and discusses major popular application areas of machine learning. Figure 9 gives the main branches of machine learning and the offshoot disciplines of each. It also depicts how different researchers have used the major ML algorithms to solve problems in the respective domains. The application areas of ML are vast, as seen by the depiction in Figure 9. Therefore, this study summarizes the application area into ten elaborate areas which are discussed below.

*i. Predictive and decision-making*

Most ML research has been carried out in this domain, where ML drives the intelligent decision-making process through data-driven predictive analytics, for instance, suspect identification, fraud detection (Hajjami et al., 2021), and many more. ML is also helpful in identifying customer preferences and behavior, production line management, scheduling optimization, and inventory management. As seen from Table 7, the keywords "prediction" and "detection" represent the third and fourth most frequently used keywords for research in ML. Nwaila et al. (2019) designed a machine learning algorithm for point-wise grade prediction and automatic facies identification based on gold assay and sedimentological data for the South African Witwatersrand Gold ores.

*ii. Cybersecurity and threat intelligence*

Cybersecurity is a cardinal area of intervention in Industry 4.0, typically protecting networks, systems, hardware, and data from digital attacks. Machine learning techniques have been used to detect security breaches through data analysis to identify patterns and detect malware or threats. The common ML technique for identifying cyber breaches is the clustering technique. Also, deep learning has been used to design security models that can be used on large-scale security datasets (MacQueen, 1967). Mbona and Eloff (2022) designed a semi-supervised machine learning approach to detect zero-day (new unknown) intrusion attacks based on the law of anomalous numbers to identify significant network features that effectively show anomalous behaviour. Similarly, Benlamine et al. (2016), used a machine learning model to evaluate emotional reactions in virtual reality environments where the face is hidden in a virtual reality headset, making facial expression detection using a webcam impossible. Several machine learning techniques have been used to identify and classify spam e-mails (Bassiouni, Ali & El-Dahshan, 2018).

*iii. Internet of Things (IoT) and smart cities*

The Internet of Things (IoT) is another vital area of the fourth Industrial revolution. The goal is to make objects smart by allowing them to transmit data and automate tasks without human interaction. Therefore, IoT is a frontier in enhancing human activities, such as smart homes, cities, agriculture, governance, healthcare, and more. Adenugba et al. (2019) proposed a machine learning-based Internet of Everything for a smart irrigation system for environmental sustainability in Africa. Their solar-powered smart irrigation system uses a machine learning radial basis function network to predict the environmental condition that controls the irrigation system.

*iv. Traffic prediction*

The economy of a city or country thrives when an efficient transport system exists. A community's economic growth comes with challenges such as high traffic volume, accidents, emergencies, high pollution, and more. Therefore, ML-driven smart city models can help predict traffic anomalies (Essien, Petrounias, Sampaio, & Sampaio, 2021). Also, ML techniques can analyze travel history data to predict possible hitches or recommend alternative routes to commuters (Boukerche & Wang, 2020)

*v. Healthcare*

Machine learning techniques have been applied in healthcare for diagnosing and prognostic diseases, omics data analysis, patient management, and more (Acharya, Hagiwara, Sudarshan, Chan, & Ng, 2018; Kim & Tagkopoulos, 2018). The Coronavirus disease (COVID-19) outbreak elicited the use of machine-learning techniques to help combat the pandemic (Kushwaha et al., 2020). Deep learning also provides exciting solutions to medical image processing problems and is a crucial technique for potential applications, particularly for the COVID-19 pandemic (Oh, Park, & Ye, 2020). Machine learning technique has also been used in Malaria incidence prediction to address the serious challenge it poses to socio-economic development in Africa (Nkiruka, Prasad & Clement, 2021). Heart failure phenotypes were clustered based on multiple clinical parameters using unsupervised machine learning techniques by Mpanya et al. (2003) to assist in diagnosing, managing, risk stratification and prognosis of heart failure. Machine learning has been deployed in predicting the present or future status of a disease or a disease's future course using machine learning and regression models (Anne-Laure Boulesteix et al.,2019). Patients can be classified based on disease risk or disease probability estimation through machine learning approaches (Kruppa, Ziegler & König, 2012). Brain MRIs can be classified for detecting brain tumors using a machine learning-based deep neural network classifier (Moshen et al.,2018). Other medical diagnoses that use machine learning include electrocardiograms (Salem, Revett & El-Dahshan, 2009) and cancer disease diagnosis (Sweilam, Tharwat & Moniem, 2010).

*vi. E-commerce*

ML techniques have been used to build systems that help businesses understand customers' preferences by analyzing their purchasing histories. These systems can recommend products to potential customers. Companies would use these systems to know where to position product adverts or offers. Many online retailers can better manage inventory and optimize logistics, such as warehousing, using predictive modeling based on machine learning techniques (Ezugwu, 2021). Furthermore, machine learning techniques enable companies to maximize profits by creating packages and content tailored to their customer's needs, allowing them to maintain existing customers while attracting new ones. Customers' creditworthiness can be determined through customers' credit scoring based on machine learning classification methods (Kruppa et al. 2013). In retail market operations, a machine learning tool has been designed to assist retailers in increasing access to essential products by improving essential product distribution in uncertain times due to the problem of panic buying (Adulyasak et al.,2020).

*vii. Natural language processing (NLP)*

NLP and sentiment analysis involve processes that could enable computer reading, understanding, and processing of spoken or written language (Otter, Medina, & Kalita, 2020). Some examples of NLP-related tasks include virtual personal assistants, chatbots, speech recognition, document description, and language or machine translation. Sentiment Analysis or Opinion Mining uses the result of NLP to mine information or trends that could translate to moods, views, and opinions from huge data collected from different social media platforms (Babu & Kanaga, 2022). For instance, politicians can use sentiment analysis to ascertain the perceived views of the electorate about their candidate.

*viii. Image, speech, and pattern recognition*

Machine learning has significant application in this domain, where different ML techniques have been used to identify or classify real-world digital images (Oyelade & Ezugwu, 2021). A typical example of image recognition includes labeling digital images from an X-ray as cancerous. Like image recognition, speech recognition deals with sound and linguistic models (Chiu et al., 2018). Finally, pattern recognition aims to identify patterns and expressions in data (Anzai, 1992). Several machine-learning techniques, such as classification, feature selection, clustering, or sequence labeling, have been used in this area.

*ix. Sustainable agriculture*

Sustainable agricultural practices help improve agricultural productivity while reducing negative environmental impacts (Adnan, Nordin, Rahman, & Noor, 2018; Sharma, Kamble, Gunasekaran, Kumar, & Kumar, 2020). Sustainable agriculture is knowledge-intensive and information-driven, where farmers make decisions based on available information and technology such as the Internet of Things (IoT), mobile technologies, and devices. Machine learning techniques are applied to predict crop yield, soil properties, irrigation requirements, weather, disease detection, weed detection, soil nutrient management, livestock management, demand estimation, production planning, inventory management, consumer analysis, and more. Machine learning techniques have been used to predict the level of insect infestation with its associated damage in maize farms (Nyabako et al., 2020). In Hengl et al. (2017), spatial predictions of soil micro and macro nutrients were carried out using machine learning techniques to support agricultural development, monitoring and intensifying soil resources. Identifying and mapping ecosystems are important in supporting food security and other important environmental indicators for biotic diversity. Tchuenté et al. (2011) developed two machine learning approaches to ecosystem mapping in the African continent-scale to classify the African ecosystem based on the Normalized Difference Vegetation Index (NDVI) dataset. Andraud et al. (2021) applied machine learning for Benthic habitat mapping to characterise seafloor substrate using geophysical data at Table Bay, southwestern South Africa. Computer vision and machine learning techniques have been used in the evaluation of food quality and the grading of crops. Seminary et al. (2014) designed machine learning techniques using feature fusion and support vector machines for classifying infected or uninfected tomato fruits based on the external surface of the tomato fruits.

*x. Pollution Control*

Air pollution is regarded as one of the world's most immense public and environmental health challenges, with its adverse effects on the ecosystem, human health, and climate. Gaps in air quality data in the middle- and lower-income countries limit the development of policies relating to air pollution control with its resultant negative health impacts due to exposure to ambient air pollution. Long-term exposure to ambient air pollution is associated with an increase in mortality rates in these countries. There is a need for accurate and reliable estimates of air pollution prediction for land use regression. Coker et al. (2021) proposed a land use regression model based on low-cost particulate matter sensors and machine learning to accurately estimate the exposure to air pollution in eastern and central Uganda – a sub-Saharan African country. The goal is to use low-cost air quality sensors in land use regression modelling to accurately predict the fine ambient particulates matter air pollution in the urban areas which will be estimated monthly. Amegah (2021) also used machine learning techniques with low-cost air quality sensors for air pollution assessment and prediction in urban Ghana. Zhang et al. (2021) developed a machine learning model using the random forest for estimating the daily fine particulate matter concentration in the industrialized Gauteng province in South Africa based on socioeconomic, satellite aerosol optical depth, meteorology and land use data.

*xi. Climate System*

In estimating global gridded net radiation and sensible and latent heat alongside their uncertainties, machine learning has been deployed to merge energy flux measurements with meteorological and remote sensing data for accurate estimation (Jung et al., 2019). The negative impact of climate change on human life informed the need for its study and prediction. Machine learning models have been employed to study the relationship between greenhouses gases emissions and climate variable change rhythm. Ibrahim, Ziedan & Ahmed (2021) explored the application of ML techniques to climate data for building an ML models for predicting climate variable states for the long and short term in North-East Africa. This is employed in climate mitigation and adaptation as well as in determining the acceptable level of greenhouse gases with their corresponding concentration to avoid climate crises and events. Sobol, Scott & Finkelstein (2019) utilized supervised machine learning to modern pollen assemblages in Southern Africa to understand biome responses to global climate change and determine specific biomes or bioregions representations. Probabilistic classification for fossil assemblages was generated for the reconstruction of past vegetation.

The continual negative effect of climate change and human-induced ecological degradation worsens the environmental pressures on human livelihoods in many regions, resulting in an increased risk of violent conflict. With reference to the African continent, Hoch et al. (2021) projected sub-national armed conflict risk along three representative concentration pathways and three shared socioeconomic pathways using machine learning methods. The role of hydro-climatic indicators in driving armed conflict was assessed. According to their report, climate change increases the projection for armed conflict risk in Northern Africa and substantial parts of Eastern Africa. The role of ML in armed conflict risk projection is to assist the policy-making process in handling climate security. To combat the adverse effect of deforestation and climate change on accurate weather information, Nyetanyane & Masinde (2020) proposed a machine learning model that uses climate data, vegetation index and indigenous knowledge to predict the onset of favourable weather seasons for crop cultivation, monitoring and prediction of crop health.

*xii. Soil Analysis*

The need for detailed soil information to assist in agricultural productivity modelling as well as to aid global estimation of the organic carbon in the soil has grown over time. Moreover, in areas affected by climate change, the need arises for spatial information about the parameters of soil waters. According to Folberth et al. (2016), obtaining accurate information about soil may be important in the prediction of the effect of climate change on food production. Hengl et al. (2017) presented an improved version of the SoilGrids system for global predictions for standard numeric soil properties, including the organic carbon, Cation Exchange Capacity, bulk density, soil texture fractions, coarse fragments and pH, as well as predicting the distribution of soil classes and depth to bedrock based on the USDA and World Reference Base classification system.

In the following paragraph, we critically discuss one of the research niche areas in which Africa has led after the United States, Canada and China, specifically in Quantum Computing machine learning research. The South Africa Quantum Technology Initiative (SA QuTI) was established in 2021 as a national undertaking that seeks to create conducive conditions for a globally competitive research environment in quantum computing technologies. Moreover, the University of KwaZulu-Natal has been leading in producing significant research output in the quantum machine learning research domain, championed by Professor Petruccione. A more detailed discussion of the quantum computing research in presented next.

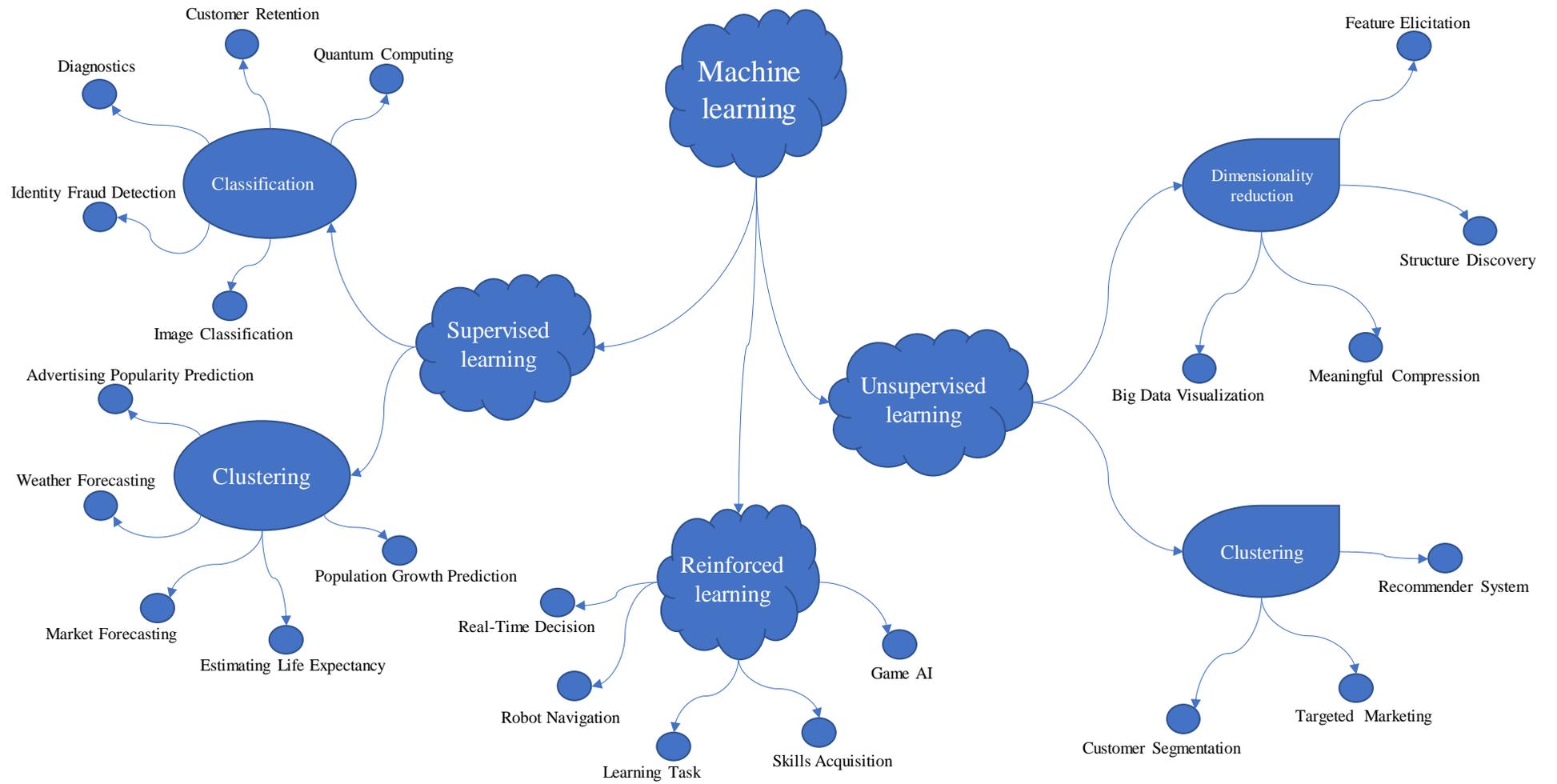

Figure 9. The main branches of machine learning and the offshoot disciplines of each

### 4.4. Quantum-Based Machine Learning Research

Another prominent research area in machine learning that has been actively engaged in Africa is the deployment of quantum computing to improve classical machine learning algorithms. Quantum computing manipulates the quantum system for information processing for a substantial computational speed. In quantum computing, the classical two states 0 and 1 of conventional computing are replaced with the superposition of qubit (quantum bit) of the two states |0⟩ and |1⟩, which allows many different computation paths simultaneously. Quantum machine learning involves the development of quantum algorithms for solving typical machine learning problems to harness the efficiency of quantum computing. The classical machine learning algorithms are adapted to run on a quantum computer. In the current era of the explosive growth of information, the adoption of quantum machine learning for various machine learning applications has been an active area of research as it is a promising area of an innovative approach to improving machine learning.

Schuld, Sinayskiy, & Petruccione (2015) presented a systematic overview of the emerging field of quantum machine learning, describing the approaches, technical details, and future quantum learning theory. The presentation included discussions on the various approaches for relating seven standard methods of the classical machine learning algorithms: support vector machine, k-nearest neighbour, neural network, k-means clustering, hidden Markov model, decision trees and Bayesian theory to quantum physics. The discussion focused mainly on the quantum machine learning approach for pattern classification and clustering

Pattern classification is one of the major tasks under supervised machine learning. Most quantum machine learning algorithms are built to address this area of machine learning to extend or improve the classical version. Schuld, Sinayskiy and Petruccione (2014) used the pattern classification examples to briefly introduce quantum machine learning. Their work presented an algorithm for quantum pattern classification using Trugenberge's proposal to measure Hamming distance on the quantum computer. Schuld, Fingerhuth and Petruccione (2017) implemented a distance-based classifier using a quantum interference circuit. In their approach, a new perspective was proposed where the distance measure of a distance-based classifier was evaluated using quantum interference in quantum parallel instead of the usual approach of the quantum machine merely mimicking the classical machine learning methods. Their approach was demonstrated on a simplified supervised pattern recognition task based on binary pattern classification.

The kernel-based machine learning method is another aspect of machine learning where quantum computing has been applied for data analysis application areas. The ability of quantum computing to efficiently manipulate exponentially large quantum space enables the fast evaluation of the kernel function more efficiently than classical computers. Blank et al. (2022) presented a compact quantum circuit for constructing a kernel-based binary classifier. Their model incorporated compact amplitude encoding of real-valued data, which reduced the number of qubits by two and linearly reduced the number of training steps. Another kernel-based quantum binary classifier was presented by Blank et al. (2020). Their distance-based quantum classifier has its kernel designed using the quantum state fidelity between the training and the test data so that the quantum kernel can be systematically tailored with a quantum circuit. The training data can be assigned arbitrary weight, and the kernel can be raised to arbitrary power.

The development of the quantum kernel method and quantum similarity-based binary classifier exploiting feature quantum Hilbert space and quantum interference brought a great opportunity for enhancing classical machine learning through quantum computing. In Park, Blank and Petruccione's (2020) work, the general theory of the quantum kernel-based classifier was extended to lay the foundation for advancing quantum-enhanced machine learning. The authors focused on using squared overlap between quantum states as the similarity measure to examine the minimal and essential ingredients for quantum binary classification. Their work also considered other extensions relating to measurement, ensemble learning and data type.

Schuld, Sinayskiy and Petruccione (2016) designed an algorithm for pattern classification with linear regression on a quantum computer. Their approach focused on solving linear regression problems from the perspective of machine learning, where new inputs are predicted based on the dataset. Their algorithm produced the same result as the least square optimisation method for classical linear regression in a logarithmic time dependent on the feature vector's number N and independent of the training dataset size if presented as quantum information.

In Schuld and Petruccione (2018), the authors introduced the quantum ensembles of quantum classifiers with parallel execution of each quantum classifier and the resulting combined decision accessed using a single qubit measurement. An exponentially large machine learning ensemble increases the performance of individual classifiers in terms of their predictive power and the ability to bypass the need for the training session. The ensemble was designed in the form of a state preparation scheme to evaluate each classifier's weight. Their proposed framework permits the exponential combination of many individual classifiers that require no training, like the classical Bayesian learning, and is credited with a quantum computing learning that is optimization-free.

In most kernel-based quantum binary classifiers, the algorithms require an expensive, repetitive procedure of quantum data encoding to estimate an expectation value for reliable operation resulting in high computational cost. Park, Blank and Petruccione (2021) proposed a robust quantum classifier that explicitly calculates the number of repetitions necessary for classification score estimation with a fixed precision to minimize the program resource overhead.

## 4.5. Renewable Energy

In renewable energy and bioprocess modelling, Kana et al. (2012) reported on the modelling and optimization of biogas production on mixed substrates of sawdust, cow dung, banana stem, rice bran and paper waste using a hybrid learning model that combines ANN and Genetic Algorithm. In another study, Whiteman and Kana (2014) investigated the relevance of ANN in modelling the relationships between several process inputs for fermentative biohydrogen production and, after that, they suggested that the ANN model is more reliable for navigating the optimization space relative to the different parameters at play for the biohydrogen production system. The authors Sewsynker et al. (2015) also reported the use of ensembles of ANNs in the modelling of biohydrogen yield in microbial electrolysis cells. The study showed that the employed ANNs model could accurately model the non-linear relationship between the physicochemical parameters of microbial electrolysis cells and hydrogen yield due to the ANNS capability to successfully navigate the optimization window in microbial electrolysis cell scale-up processes. ML has been used for multi-objective intelligent energy management for the microgrid to improve efficiency in microgrid operation (Chaouachi et al., 2012). A hybrid ML technique has been used for predicting solar radiation based on meteorological data (Olatomiwa et al., 2015) with an analysis of the influence of weather conditions in different regions of Nigeria. A machine learning model for predicting the daily global solar radiation was designed in Morocco by Chaibi et al. (2021).

## 4.6. Prospects, challenges, and recommendations

The prospects of ML research in Africa are enormous. It also has challenges, such as bioinformatics research in Africa being limited by the availability of diverse and high-volume biomedical data for accurate analysis (Mulder et al., 2017). As data is central to ML, the Human Heredity & Health in Africa (H3Africa) consortium is championing efforts at generating and publicly publishing large genomics datasets of Africans (Rotimi et al., 2014). Another obstacle is the lack of a computing backbone which includes internet connectivity and cloud computing, which leads to data outsourcing to the developed world (Nordling, 2018)

Similarly, the prospects of ML will be inactive if appropriate investments in this direction are not made. Also, teaching AI techniques, including ML, must be improved and sustained. An adequate legal framework must be in place to ensure ethical research and innovative development (Novitske, 2018). A framework for support and collaboration with foreign agencies must be encouraged. For instance, the strategic partnership between the Smart Africa alliance and the German Ministry for Economic Cooperation and Development aims to support Africa's development through digital innovations (Smart Africa, 2021; Digital Africa, 2021).

The diverse applicability and techniques promoting the use of AI systems have received more research efforts from ML. The increasing use of ML algorithms and their subsidiary methods, such as DL, has further shown the computational power of CNN, RNN, LSTM and hybrid models. These models have demonstrated outstanding performance in pattern recognition, classification, feature extraction, segmentation and other learning approaches. Interestingly, while current studies and state-of-the-art are majoring in hybridizing sequence models such as RNN with pixel models such as CNN for multimodal computation, little is mentioned on machine reasoning. The descent of machine reasoning from the aspect of knowledge representation and reasoning may not be directly associated with machine learning. Still, the successful integration of these two branches of AI holds the possibility for achieving high-performing systems in the near future. Machine learning, on the one hand, allows for fine-tuning models and their parameters in a manner that sets those parameters to enable the machine to behave in a manner simulated by a human.

On the other hand, machine reasoning provides means for formalising the existing body of knowledge siloed away in legacy systems for achieving reasoning and inference. Combining these two aspects of machine automation will promote what is termed neuro-symbolic systems, which allows for neural networks and rules with formalized knowledge to interface in a manner to drive new state-of-the-art AI applications. We motivate for redirection of study in AI, ML, and DL among African researchers to consider this aspect of learning and reasoning.

Another prospective integration of branches of AI which promises to promote the discovery of super intelligent systems is the application of clustering and optimization methods to the models of DL and deep reinforcement learning (DRL). Research in the design of DRL models is now yielding and controlling self-driving cars, fully automated systems, robotics and other aspects of autonomous systems. Although DRL draws from the concept of DL, we consider that identifying some features in DL models (e.g. CNN, RNN, LSTM, GRU and their hybrids) and effectively integrating them with DRL will uncover some outstanding high-level performance with regards to machine intelligence. Researchers in Africa are likely to develop an interesting outcome in this aspect, considering their progress in using these models in their current isolated form of use. Moreover, clustering and metaheuristic methods promise to provide relevant and hardcore optimization solutions to improve the integration of the hybrids mentioned earlier in this paragraph. Of course, we have seen several usages of metaheuristic methods in DL models and with the increasing use of clustering methods. This study motivates a way forward for an in-depth look into the possible interfacing of DRL, DL and some clustering methods with the use of optimization techniques for bolstering performance and computational cost.

The applicability of the resulting intelligent systems from the current and future state-of-the-art in AI, ML and DL is still in its infancy stage in Africa. The COVID-19 pandemic demonstrated that Africa still lags behind in adopting some of the research outcomes from its researchers. Although the effect of the pandemic is considered not to be very destabilizing when compared with other continents, the lesson that must be learnt is that Africa must prepare for a future pandemic by leveraging on the research outcome coming from research centres in Africa. Therefore, this holds prospects and challenges that can spur on or open up new interesting research areas. For instance, consider applying ML methods to building smart cities across Africa. This will draw from significant AI methods and systems successfully designed and developed for smarting out all infrastructures and facilities in such cities. Consider also the application of research efforts in Computer Vision to the challenge of aiding Africa's transport and communication (T&C) system. Firstly, the pedestrian system must be automated and integrated with the T&C system for an effective AI-driven computing network. We advocate for state-sponsored research in this direction as it holds the prospect of improving road connectivity and trade across the continent. Another interesting aspect of AI's applicability to Africa's peculiarities is in the area of crime monitoring and surveillance. For the latter, the progress made in Computer Vision combined with the Internet of Things (IoTs) has already provided for the deployment of facilities to aid the state's surveillance system and the law enforcement commissions. The former crime detection and monitoring concept will benefit from recent deep learning-driven natural language processing (NLP) methods to analyze a pool of data floating on different social media platforms and other text-driven systems for effective crime detection. Motivated by the increasing hosting of deep learning indaba conferences in Nigeria, Tunisia and South Africa, with most of them promoting DL-NLP, there is now a greater prospect of the application of these methods to crime detection and monitoring. In addition to this, this DL-NLP method showed that the rich multi-lingual formation across all tribes and peoples in Africa could interact more effectively and develop information-sharing mechanisms through the use of machine translation. For instance, it is well known that peoples speak languages like Hausa, Swahili, Yoruba, Arabic, and isiZulu in different countries. The adoption of machine translation will therefore help to build on this communication skill and close gaps. Lastly, with the plethora of research outcomes in medical image analysis and AI-driven computer-aided diagnosis (CAD) systems, healthcare delivery and medical sciences will receive a boost in health centres across Africa.

A current challenge which needs to be addressed to promote research in ML in Africa is an intensive and intentional investment in computational infrastructure. ML and DL experiments demand high computational power with the requirement for memory and graphical processing units (GPU), and reliable power grids. Stakeholders and government must integrate their thinking and resources to build a cohesive and robust computational infrastructure to help support researchers' efforts during experimentation and deployment. This is necessary to allow for rigorous testing and experimentation of new models capable of becoming new state-of-the-art globally. Moreover, the sustenance of startup hubs, as seen in Morocco, Nigeria, Ghana, Kenya and South Africa, needs to be promoted to allow for the convergence of test hubs for AI solutions being developed by African youths.

## 5. Conclusions

Machine learning evolved as a branch in AI, focusing on designing computational methods and learning algorithms that model humans' natural learning patterns to address real-life problems where human capability is limited or restricted. This paper presents a background study of ML and its evolution from AI through ML to DL, elaborating on the various categories of learning techniques (supervised, unsupervised, semi-supervised and reinforcement learning) that have evolved over the years. It also presents the contribution of different ML researchers across major African universities from niche areas or multi-disciplinary domains.

Moreover, a bibliometric study of machine learning research in Africa is presented. In total, 2761 machine learning-related documents, of which 89% were articles with at least 482 citations, were published in 903 journals in the Science Citation Index EXPANDED from 54 African countries between 1993 and 2021. There are 12 topmost frequently cited documents, of which five were review articles. Significant interest in machine learning research in Africa began in 2010, with the number of articles increasing slightly from 14 to 98 in 2017 and which then increased with a huge leap to 1035 articles by 2021. The highest article citation was recorded in 2013. The top four productive categories in the Web of Science, where more than 100 articles were published, include "electrical and electronic engineering", "information systems computer science", "artificial intelligence computer science", and "telecommunication", each recording 20%, 18%, 14% and 12% of the total number of articles respectively. The most productive journal is IEEE Access, with 192 articles (7.8%).

The top five journals with IF2021 of more than 60 published six of the articles: World Psychiatry (2), Nature (1), Nature Energy (1), Nature reviews (1) and Science (1). International collaborative articles recorded the highest number of articles, 74% involving 43 African countries and 103 non-African countries, while the remaining single-country articles were from 16 African countries. Egypt dominated with 31% of the total article publication, 29% being single-country articles and 32% being internationally collaboratively published. Ten African countries had no publication in machine learning-related articles, while 64% of the remaining countries had no single-country articles. Egypt and South Africa had similar development trends, but Egypt recorded a noticeably sharp increase in the last three years. Cairo University in Egypt ranked top among the most productive African institutions, with the University of

Kwazulu-Natal in South Africa ranking top in three of the six publication indicators. King Saud University in Saudi Arabia tops the list of the five non-African institutions with 30 or more inter-institutionally collaborative articles with Africa.

Among the top ten most frequently cited machine learning-related articles in Africa, authors published five from Egypt, followed by authors from South Africa with two articles. The most cited article was published by Wright and Ziegler in 2017 from the University of Lubeck in Germany and the University of KwaZulu-Natal in South Africa, while Adadi and Berrada published the article with the most impact in the recent year 2021 by Sidi Mohammed Ben Abdellah University in Morocco in 2018. The four top keywords used by authors in African machine learning-related articles are classification, deep learning, feature extraction and random forest.

Furthermore, a review of machine learning techniques and their applications in Africa in recent years was presented, identifying the main branches of ML and their offshoot disciplines. The nine most significant machine-learning application areas in Africa were identified and discussed. Research on quantum implementations of machine learning algorithms in Africa for performance improvement of the classical machine learning techniques was also reviewed. Moreover, quantum machine learning is one area of interest in ML research which has positively projected the image of African research scholars from the University of KwaZulu-Natal and has equally attracted global attention from quantum computing enthusiasts. Finally, the prospects and challenges with recommendations regarding ML research in Africa were discussed in detail.

**Acknowledgement**

NA.

**Ethical Approval**

NA

**Competing interests**

The authors declare that there is no conflict of interest with regard to the publication of this paper.

**Funding**

NA

**Availability of data and materials**

All data generated or analyzed during this study are included in this article.